\documentclass[a4paper,11pt]{article}
\pdfoutput=1 

\usepackage{jheppub} 

\usepackage[T1]{fontenc} 

\title{\boldmath One-loop impact factors for heavy quarkonium production: $S$-wave case}

 \author{Maxim A. Nefedov}
 \affiliation{Universit\'e Paris-Saclay, CNRS, IJCLab, 91405 Orsay, France}

\emailAdd{maxim.nefedov@ijclab.in2p3.fr}

\abstract{With the aim to extend the study of inclusive heavy quarkonium production at forward rapidities with the resummation of high partonic center-of-momentum-energy logarithms beyond Leading Logarithmic Approximation (LLA), the explicit analytic results for one-loop corrections to the following impact factors had been obtained: $\gamma R \to Q\bar{Q}\left[^1S_0^{[8]} \right]$, $g R \to Q\bar{Q}\left[^1S_0^{[1]} \right]$, $gR\to Q\bar{Q}\left[ ^1S_0^{[8]} \right]$ and $gR\to Q\bar{Q}\left[ ^3S_1^{[8]}\right]$, with $R$ being the Reggeized gluon and $Q$ is the heavy quark. The computation is done in the framework of Lipatov's gauge-invariant EFT for Multi-Regge processes in QCD with the tilted-Wilson-line regularisation for rapidity divergences.  As expected, only single-logarithmic rapidity divergence proportional to the one-loop Regge trajectory of a gluon remains in the final result for impact-factors. Numerical comparison with Regge limits ($s/(-t) \gg 1$) of one-loop QCD amplitudes, described in the paper, provides a strong cross-check of obtained results. The relations of obtained results with other regularisation schemes for rapidity divergences used in low-$x$ physics, such as BFKL scheme, High-Energy Factorisation (HEF) scheme and shockwave scheme, are given.}

\begin{document} 
\maketitle
\flushbottom

\section{Introduction}

Understanding of hadronisation mechanism of heavy quarks ($Q$) into quarkonia (charmonia or bottomonia) is one of the most important open problems in QCD, see recent reviews~\cite{Brambilla:2010cs,Brambilla:2014jmp,Lansberg:2019adr,Arbuzov:2020cqg,Chapon:2020heu}. The relative (squared) velocity ($v^2$) of heavy quarks in these bound states is estimated in potential models to be $v^2\sim 0.3$ for charmonia and $\sim 0.1$ for bottomonia, suggesting that genuine many-body QCD effects in the hadronisation process can be systematically taken into account as corrections in $v^2$, on the same footing with relativistic effects. The non-relativistic QCD (NRQCD) factorisation formalism~\cite{Bodwin:1994jh} for heavy quarkonium production implements the expansion of cross section in $v^2$, representing it as a sum of contributions of various intermediate Fock-states of $Q\bar{Q}$-pair with quantum numbers ${}^{2S+1}L_J^{[1,8]}$ with $S$ being the total spin of the pair, $L$ -- the orbital momentum and $J$ -- total angular momentum, while labels $^{[1,8]}$ denote colour-singlet or colour-octet states of the $Q\bar{Q}$, with contributions of colour-octet intermediate states, as well as states with higher orbital momentum, being suppressed by $v^2$.   

Recently, the problem of perturbative instability of NRQCD-factorisation calculations at high hadronic or photon-hadron collision energies have been addressed on  examples of total cross sections of hadroproduction of $\eta_c$-mesons~\cite{Lansberg:2021vie} or photoproduction of $J/\psi$ mesons~\cite{Lansberg:2023kzf} in the LO in $v^2$ approximation, where the colour-singlet state of $Q\bar{Q}$ dominates the expansion. As shown in Refs.~\cite{Lansberg:2021vie,Lansberg:2023kzf}, to obtain physically reasonable predictions for cross sections integrated over transverse momentum of a quarkonium ($p_T$), one has to supplement the NLO QCD computation in  Collinear Factorisation(CF) with the resummation of the logarithms of {\it partonic} center-of-mass energy ($\hat{s}$), which in the Leading-Logarithmic-Approximation (LLA) corresponds to the resummation of the series of higher-order corrections $\propto\alpha_s^n \ln^{n-1}(\hat{s}/M^2)$ (with $M$ being the quarkonium mass) to the partonic coefficient functions ($d\hat{\sigma}$). This resummation is done with the help of High-Energy Factorisation (HEF) formalism~\cite{Catani:1990xk,Catani:1990eg,Collins:1991ty,Catani:1994sq}. In Refs.~\cite{Lansberg:2021vie,Lansberg:2023kzf} the usage of the subset of the LLA -- the {\it Doubly Logarithmic Approximation (DLA)} of the HEF is advocated as a proper starting approximation if the usual fixed-order Parton Distribution Functions(PDFs) are to be used in the calculation. 

\begin{figure}
    \centering
    \includegraphics[width=0.7\linewidth]{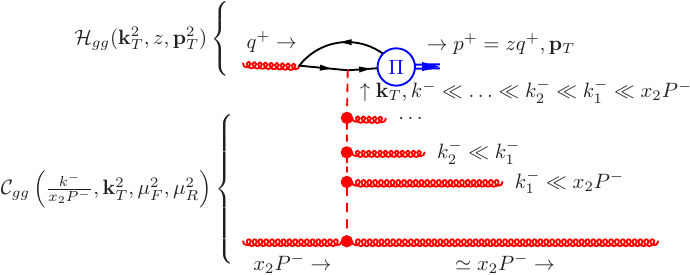}
    \caption{Typical Feynman diagram contributing to the coefficient function of CF $d\hat{\sigma}_{gg}$ in the LLA in $\ln \hat{s}/M^2$ for the forward production of the NRQCD intermediate states $Q\bar{Q}[{}^1S_0^{[1,8]},{}^3S_1^{[8]},{}^3P_J^{[1,8]}]$. Dashed lines denote Reggeised gluons. The $q^+=x_1P^+$ and the LO impact-factor ${\cal H}_{gg} \propto \delta({\bf k}_T^2-{\bf p}_T^2) \delta(1-z)$ as in Eq. (\ref{eq:H-LO-def}) below. The $\Pi$-blob denotes the spin and colour projectors on the corresponding NRQCD intermediate state of the $Q\bar{Q}$, discussed in Sec.~\ref{sec:LO}.  }
    \label{fig:HEF-schematic}
\end{figure}

It is necessary to go beyond of the DLA, used in Refs.~\cite{Lansberg:2021vie,Lansberg:2023kzf}, to improve the accuracy of these computations, in particular to reduce the scale uncertainty. Also, in order to go beyond the $p_T$-integrated cross sections, which are difficult to precisely measure in experiment, one can consider $p_T$-differential observables for $p_T\sim M$. These observables also will be sensitive to the large logarithms of $\hat{s}$ as long as the hadronic or photon-hadron collision energy ($S=P^+P^-$) is sufficiently large to guarantee that the region of $\hat{s} \gg M^2$ contributes significantly. Therefore a reasonable continuation of the research program started in Refs.~\cite{Lansberg:2021vie,Lansberg:2023kzf} is to consider the $p_T$-differential cross sections of hadro or photo production of quarkonia at forward rapidities, which are characterised by a very asymmetric LO CF  kinematics of initial-state partons, with one parton carrying the momentum fraction $x_1\lesssim 1$ and another one having $x_2\ll 1$. From the low-$x$ parton ``side'' of the process, the HEF is applicable and the corresponding factorisation of the partonic coefficient function $d\hat{\sigma}$ in terms of the HEF coefficient function (${\cal H}({\bf k}_T)$) and the HEF resummation factor (${\cal C}_{gi}$), defined in Refs.~\cite{Lansberg:2021vie,Lansberg:2023kzf}, takes place:
\begin{equation}
    \frac{d\hat{\sigma}_{ij}}{dz d{\bf p}_T^2} = \frac{1}{2M^2} \int \frac{d^2{\bf k}_T}{\pi} {\cal C}_{ig} \Bigl(\frac{\hat{s}}{M^2},{\bf k}_T^2,\mu_F,\mu_R \Bigr) {\cal H}_{gj}({\bf k}_T^2,z,{\bf p}_T^2)+O\left( \frac{M^2}{\hat{s}},\frac{{\bf p}_T^2}{\hat{s}},\text{NNLLA}\right), \label{eq:sighat-factor}
\end{equation}
where $z$ is the fraction of momentum of the moderate-$x$ parton (one with $x_1\lesssim 1$), which is carried-away by the produced quarkonium and ${\bf p}_T$ is it's transverse momentum. The diagrammatic sketch of this factorisation is also shown in the Fig.~\ref{fig:HEF-schematic} for the LLA case.  As indicated, this factorisation of the coefficient function is valid up to power corrections suppressed at large $\hat{s}$ and up to Next-to-Next-to-LLA (NNLLA) corrections $\propto \alpha_s^{n+2} \ln \hat{s}/M^2$, which can not be put in the factorised form of Eq.~(\ref{eq:sighat-factor}) due to Multi-Reggeon exchanges starting to contribute at this order~\cite{DelDuca:2001gu,Fadin:2017nka,Caron-Huot:2017fxr}, so Eq.~(\ref{eq:sighat-factor}) has to be generalised to go beyond NLLA.

In the present paper the one-loop corrections to the coefficient function ${\cal H}_{gj}({\bf k}_T^2,z,{\bf p}_T^2)$ in Eq.~(\ref{eq:sighat-factor}), also known as {\it impact factor} in BFKL~\cite{BFKL1,BFKL2,BFKL3} terminology, are computed for the production of several $S$-wave states of a $Q\bar{Q}$-pair, relevant for NRQCD-factorisation computations: ${}^1S_0^{[1]}$, ${}^1S_0^{[8]}$ and ${}^3S_1^{[8]}$. The same computation for the $P$-wave states  is currently in progress. 

Besides computations of inclusive quarkonium production cross sections in HEF, results derived in the present paper can be also used for the computation of heavy quarkonium pair or quarkonium+jet pair hadroproduction cross sections at large rapidity separation between components of the pair. Computations of this type are currently done only in the LLA~\cite{He:di-Jpsi} or in the NLLA with fragmentation approximation for the quarkonium impact-factor~\cite{Celiberto:2022dyf}, which is not applicable at moderate $p_T\sim M$. 

The present paper has the following structure. In Sec.~\ref{sec:EFT} the main tool used in the computation, which provides us with the operator definition of a Reggeised gluon -- the Lipatov's EFT for Multi-Regge processes in QCD~\cite{Lipatov95} is described\footnote{The operator definition of Ref.~\cite{Caron-Huot:2013fea} is equivalent to the definition provided by the EFT in the LLA and NLLA, while discrepancy between them is possible starting from NNLLA. However this topic is beyond the scope of the present paper. }. In Sec.~\ref{sec:diagrams}, the classes of Feynman diagrams in the EFT, contributing to quantities under consideration, are described and the algorithm of reducing them down to the small set of scalar one-loop master integrals is presented. Sec.~\ref{sec:master-ints} deals with computation of those master integrals which contain eikonal (linear) denominators together with massive quadratic propagators. In Sec.~\ref{sec:results} the results of the computation of virtual corrections to the impact-factors of processes under consideration are presented. Obtained results for finite parts of the one-loop impact factors are extensively tested by comparing them with extractions of these quantities from high-energy (Regge) limits for amplitudes $g+g\to Q\bar{Q}[{}^1S_0^{[1]},{}^1S_0^{[8]},{}^3S_1^{[8]}] + g$ or $\gamma+g\to Q\bar{Q}[{}^1S_0^{[8]}] + g$. Numerical results for these $2\to 2$ amplitudes are obtained with the help of \texttt{FORMCalc}~\cite{Hahn:2000jm}. The latter computation is nontrivial by itself due to Coulomb divergence, which arises in the considered amplitudes, and the method to deal with it is described in the Appendix~\ref{appendix:FormCalc}.  In Sec.~\ref{sec:rapidity-schemes} the relations between results for one-loop corrections to impact-factors obtained in the present paper in the tilted-Wilson-line regularisation scheme with impact factors for BFKL-type resummation calculations~\cite{Celiberto:2022dyf} (Sec.~\ref{sec:BFKL-scheme}), HEF (Sec.~\ref{sec:HEF-scheme}) or the calculations done in the scheme where the cut on the {\it projectile} light-cone component of the loop momentum regularises the rapidity divergence (the ``shockwave'' scheme, Sec.~\ref{sec:Shockwave-scheme}) are presented.  Finally, the Sec.~\ref{sec:conclusions} summarises conclusions and outlook of the present study. 

All the expressions for one-loop corrections, listed in Sec.~\ref{sec:results} are also provided in the computer-readable plain-text format in the auxiliary files included into the ArXiv submission of this paper.

\section{Gauge-invariant EFT for multi-Regge processes in QCD}
\label{sec:EFT}
The gauge-invariant EFT for Multi-Regge processes in QCD~\cite{Lipatov95} deals with processes where several groups (clusters) of QCD partons are produced in a collision of high-energy QCD partons, and the separation in rapidity between final-state clusters is assumed to be large. To derive the amplitude of such process up to corrections power-suppressed by the energy of the collision($s$) and by $e^{-\Delta y_{ij}}$ where $\Delta y_{ij}=y_i-y_j$ is the rapidity separation between clusters $i$ and $j$, the following effective Lagrangian is introduced:
\begin{equation}
  L_{\rm eff}= 4{\rm tr}\left[ R_+ \partial_T^2 R_- \right] + \sum\limits_i \left[ L_{\rm QCD}(A_\mu^{(i)}, \psi_q^{(i)}) + L_{\rm Rg}(A_\mu^{(i)}, R_+, R_-)   \right],   \label{Eq:LEFT}
 \end{equation}
where the index $i$ enumerates the rapidity-clusters and a separate copy of the usual QCD Lagrangian $L_{\rm QCD}$ is introduced for each of them, depending on it's own gluon ($A_\mu^{(i)}$) and quark ($\psi_q^{(i)}$) fields.  The interaction between clusters is carried by the fields of Reggeised gluons ($R_{\pm}$), which are  scalar fields in the adjoined representation of the colour group $SU(N_c)$ in this EFT. The kinetic part of the Lagrangian (\ref{Eq:LEFT}) leads to bare propagators, connecting the $R_+$-field with $R_-$: $i/(2{\bf k}_T^2)$, where ${\bf k}_T$ is the (Euclidean) transverse part of the momentum of the Reggeon. In the leading-power approximation w.r.t. $s\to \infty$ and $\Delta y_{ij}\to\infty$ the light-cone components of momenta of particles\footnote{$k_\pm= k^\pm = k^0 \pm k^3$ in the center-of-momentum frame of the collision.}, belonging to each cluster, are strongly ordered between clusters, e.g.: $\ldots \gg l_+^{(i-1)} \gg q_+^{(i)} \gg p_+^{(i+1)} \gg \ldots$ and  $\ldots \ll l_-^{(i-1)} \ll q_-^{(i)} \ll p_-^{(i+1)} \ll \ldots$ for momenta $l^{(i-1)}$, $q^{(i)}$ and $p^{(i+1)}$ of any particles belonging respectively to the clusters $i-1$, $i$ and $i+1$. This kinematic situation is often called \textit{(quasi-)Multi-Regge Kinematics (MRK)}. Due to the MRK, fields $R_{\pm}$ are subject to the following constraints:
 \begin{eqnarray}
 \partial_+ R_- = \partial_- R_+ = 0, \label{Eq:kin-constr-R}
  \end{eqnarray}
where $\partial_\pm=n_{\pm}^\mu\partial_\mu=2\partial/\partial x_{\mp}$ with $n_{\pm}^\mu = (1,0,0,\mp 1)^\mu$ in the center-of-momentum frame.

The ``tree-level'' interactions of fields fields $R_{\pm}$ with the QCD gluons $A^{(i)}_\mu$ in each cluster are determined by the requirement of gauge-invariant factorisation of tree-level amplitudes in the MRK up to various $i\eta$ prescriptions for the eikonal denominators $\sim 1/k_{\pm}$ which arise in these interactions. The pole prescriptions become important at loop level and can be determined from the assumption of factorisation of loop-level amplitudes in terms of Reggeised-gluon exchanges and negative signature of the one-Reggeon exchange contributions~\cite{MH_PolePrescr}. The Hermitian form of the effective action~\cite{RevLipatov97,BondZubkov} satisfies requirements of definite signature and factorisation, of the Ref.~\cite{MH_PolePrescr}:
\begin{equation}
  L_{Rg}(x)=\frac{i}{g_s}{\rm tr}\left[R_+(x) \partial_T^2 \partial_- \left(W_x[A_-]-W_x^\dagger[A_-]\right) + R_-(x) \partial_T^2\partial_+ \left(W_x[A_+]-W_x^\dagger[A_+]\right) \right], \label{Eq:L-Rg}
\end{equation}
where $W_x[A_{\pm}]$ are the past-pointing half-infinite Wilson lines, stretching in the $(+)$ or $(-)$ light-cone direction from the point $x$:
\begin{eqnarray}
  W_x[A_{\pm}]&=& P\exp\left[\frac{-ig_s}{2} \int\limits_{-\infty}^{x_{\mp}} dx'_{\mp} A_{\pm}\left(x_{\pm}, x'_{\mp}, {\bf x}_{T}\right)  \right] \nonumber \\
  &=& 1-ig_s\left(\partial_\pm^{-1}A_{\pm} \right) + (-ig_s)^2\left(\partial_\pm^{-1}A_{\pm}\partial_\pm^{-1}A_{\pm}\right)+\ldots , \label{Eq:WL-def}
  \end{eqnarray}
with $g_s$ being the coupling constant of QCD and the operators $\partial^{-1}_{\pm}$ are defined by $\partial^{-1}_{\pm}f(x)=\int\limits_{-\infty}^{x^{\mp}} dx'_{\mp}/2\ f(x_{\pm},x'_{\mp},{\bf x}_T)$, corresponding to eikonal denominators $-i/(p_{\pm}+i\eta)$ on the level of Feynman rules.  \\

Upon expansion in $g_s$, the Lagrangian (\ref{Eq:L-Rg}) generates an infinite series of {\it induced vertices} of interaction of a Reggeised gluon with $n$ QCD gluons. The simplest of them are the $R_-g$-transition vertex and $R_-gg$ interaction vertices, corresponding respectively to $O(g_s^0)$ and $O(g_s^1)$ terms in (\ref{Eq:L-Rg}):
\begin{eqnarray}
 \Delta_{-\mu_1}^{ab_1}(k,l_1)&=&(i{\bf k}_T^2)n^+_{\mu_1}\delta_{ab_1}, \label{eq:Rg-vertex-EFT} \\
 \Delta_{-\mu_1 \mu_2}^{ab_1 b_2}(k,l_1, l_2) &=&  -g_s {\bf k}_T^2 (n^+_{\mu_1} n^+_{\mu_2}) \frac{f^{ab_1b_2}}{[l_1^+]}, \label{eq:Rgg-vertex-EFT}
\end{eqnarray} 
where $k$ is the (incoming) momentum of the Reggeised gluon with the colour index $a$, $l_{1,2}$ are the (incoming) momenta of QCD gluons coupled to the vertex, with their colour and Lorentz indices denoted as $b_{1,2}$ and $\mu_{1,2}$ respectively, $l_1^+ + l_2^+=0$ due to the MRK constraint (\ref{Eq:kin-constr-R}) and the principal value prescription for the eikonal pole is denoted as
\begin{equation} 
\frac{1}{[l_+]} = \frac{1}{2}\left[ \frac{1}{l_+ + i\eta} + \frac{1}{l_+ - i\eta} \right].
\end{equation}
The vertices of interactions of $R_+$ with gluons are the same up to the flipping of $+\leftrightarrow -$ labels.

In the present paper the $R_-ggg$ and $R_-gggg$ vertices will appear, which according to the Lagrangian~(\ref{Eq:L-Rg}) have the form:
\begin{eqnarray}
&&\hspace{-5mm}\Delta_{-\mu_1\mu_2\mu_3}^{ab_1b_2b_3}=ig_s^2{\bf k}_T^2 (n_{\mu_1}^+ n_{\mu_2}^+ n_{\mu_3}^+) \sum\limits_{(i_1,i_2,i_3)\in S_3} \frac{{\rm tr}\left[T^a \left(T^{b_{i_1}}T^{b_{i_2}} T^{b_{i_3}} +  T^{b_{i_3}}T^{b_{i_2}} T^{b_{i_1}}\right) \right]}{(l_{i_3}^+ +i\eta)(l_{i_3}^+ + l_{i_2}^++i\eta)}, \label{Eq:R+_g3-vert} \\
&&\hspace{-5mm}\Delta_{-\mu_1\mu_2\mu_3\mu_4}^{ab_1b_2b_3b_4}=ig_s^3{\bf k}_T^2 (n_{\mu_1}^+ n_{\mu_2}^+ n_{\mu_3}^+ n_{\mu_4}^+) \nonumber \\
&& \times \sum\limits_{(i_1,i_2,i_3,i_4)\in S_4} \frac{{\rm tr}\left[T^a \left(T^{b_{i_1}}T^{b_{i_2}} T^{b_{i_3}} T^{b_{i_4}} - T^{b_{i_4}} T^{b_{i_3}}T^{b_{i_2}} T^{b_{i_1}}\right) \right]}{(l_{i_4}^+ +i\eta)(l_{i_4}^+ + l_{i_3}^+ +i\eta) (l_{i_4}^+ + l_{i_3}^+ + l_{i_2}^++i\eta)}, \label{Eq:R+_g4-vert}
\end{eqnarray} 
where $l_1,\ldots, l_4$ are incoming gluon momenta with $l_1^+ + l_2^+ + l_3^+=0 $ in Eq.~(\ref{Eq:R+_g3-vert}) and $l_1^++l_2^++l_3^++l_4^+=0$ in Eq.~(\ref{Eq:R+_g4-vert}), while summations are carried over permutations of three ($S_3$) or four ($S_4$) indices respectively. 

 Clearly, the phase-space integrations, as well as loop integrations over momenta of QCD partons ($l_i$) can violate the MRK, leading to {\it rapidity-divergences}, which have to be properly regularised. Following Refs.~\cite{Chachamis:2012cc, Chachamis:2012gh, Chachamis:2013hma} the covariant or ``tilted-Wilson-line'' regularisation prescription is used in the present paper, which boils down to the following replacements in all formulas above:
 \begin{equation}
     n_\mu^{\pm} \to \tilde{n}_\mu^{\pm} = n^{\pm}_\mu + r n_\mu^{\mp},\ l_{\pm} \to \tilde{l}_{\pm}=l_{\pm} + r l_{\mp},
 \end{equation}
where $0<r\ll 1$ is the regularisation parameter, and the kinematic constraint (\ref{Eq:kin-constr-R}) should be modified, to preserve gauge-invariance of the EFT for $r\neq 0$ as~\cite{Nefedov:2019mrg}:
\begin{equation}
    (\tilde{n}_{\pm} \cdot \partial) R_{\mp} =0,
\end{equation}
meaning that e.g. for $R_-(k)$ the $\tilde{k}_+=k_++rk_-=0$. Then, rapidity divergences manifest themselves as singular terms in the limit $r\to 0$, which should be taken before expansion of the amplitude in $\epsilon=(4-d)/2$ with $d$ being the dimension of space-time. Only single-logarithmic rapidity divergence $\sim\ln r$ is consistent with gluon Reggeisation and therefore all power-law and double-logarithmic terms which arise at one loop in various contributions, should cancel.

\section{Feynman diagrams and reduction to master integrals}
\label{sec:diagrams}
\subsection{Leading order amplitudes}
\label{sec:LO}
The aim of the present paper is to compute the one-loop corrections to impact-factors of the following processes:
\begin{eqnarray}
    \gamma(q)+R_-(k)&\to & Q\bar{Q}\left[ ^1S_0^{[8]} \right](p), \label{proc:ga+R->1S08}\\
    g(q)+R_-(k) &\to & Q\bar{Q}\left[ ^1S_0^{[1]} \right](p),  \label{proc:g+R->1S01} \\
    g(q)+R_-(k) &\to & Q\bar{Q}\left[ ^1S_0^{[8]} \right](p),  \label{proc:g+R->1S08} \\
    g(q)+R_-(k) &\to & Q\bar{Q}\left[ ^3S_1^{[8]} \right](p),  \label{proc:g+R->3S18}
\end{eqnarray}
with momenta $q$ ($q^2=0$, $q_-=0$, $q_+ > 0$) and $k$ ($k^2=-{\bf k}_T^2$,  $k_+=0$, $k_-> 0$), and the mass of the heavy quark pair $p^2=M^2=(2m_Q)^2$ being equal to the mass of the quarkonium ($M$) in the LO in $v^2$. 

In the LO, the HEF coefficient function of Eq.~(\ref{eq:sighat-factor}) is related with squared amplitudes\footnote{Or rather amputated Green's functions, since the Reggeon is always off-shell.} of processes (\ref{proc:ga+R->1S08}) -- (\ref{proc:g+R->3S18}) as follows:
\begin{equation}
    {\cal H}_{gj}^{\text{(LO)}} =\frac{2\pi M^2 }{(M^2+{\bf k}_T^2)^2} \left( \frac{k_-}{2} \right)^2 \frac{|{\cal M}(j(q)+R_-(k)\to Q\bar{Q}(p))|^2}{(N_c^2-1) {\bf k}_T^2} \delta(1-z)\delta({\bf k}_T^2-{\bf p}_T^2), \label{eq:H-LO-def}
\end{equation}
where $k_-=(M^2+{\bf k}_T^2)/q_+$ due to momentum conservation in the $2\to 1$ process. For the interference term between Born and one-loop corrected impact-factors, the relation is the same, with the replacement $|{\cal M}_{\rm LO}|^2 \to 2{\rm Re}[{\cal M}_{\text{1-loop}}{\cal M}^*_{\rm LO}]$.

The standard procedure of computation of short-distance coefficients of production of $Q\bar{Q}$-pair in the NRQCD intermediate state, traditionally labelled as ${}^{2S+1}L_J^{[1,8]}$, with $S$, $L$ and $J$ being the spin, orbital and total angular momentum of the $Q\bar{Q}$ and labels $^{[1,8]}$ denoting the colour-singlet(CS) or colour-octet (CO) state of the pair, prescribes for the $L=0$ case to assign the momenta of heavy quarks to be equal to the half of the momentum of quarkonium: $p_{Q}=p_{\bar{Q}}=p/2$, i.e. their relative momentum is set to zero. The desired spin state can be conveniently projected-out with the replacements for heavy quark spinors in the amplitude:
\begin{eqnarray}
    \bar{u}(p_Q) {\cal M} v(p_{\bar{Q}}) = {\rm tr} \left[{\cal M} v(p_{\bar{Q}}) \bar{u}(p_Q) \right] \to {\rm tr} \left[{\cal M} \Pi_S \right], 
\end{eqnarray}
with the well-known (see e.g.~\cite{Mangano:1996kg}) covariant projectors for $S=0$ and $S=1$ states:
\begin{eqnarray}
    \Pi_0&=&\frac{1}{\sqrt{M^3}} \left(\frac{p}{2} - \frac{M}{2} \right) \gamma_5 \left(\frac{p}{2} + \frac{M}{2}\right), \label{eq:SpinProj-0} \\
    \Pi_1&=& \frac{\varepsilon^*_\mu(p)}{\sqrt{M^3}} \left(\frac{p}{2} - \frac{M}{2} \right) \gamma^\mu \left(\frac{p}{2} + \frac{M}{2}\right), \label{eq:SpinProj-1}
\end{eqnarray}
where $\varepsilon^\mu(p)$ is the polarisation vector of the $S=1$ state. To define the projector (\ref{eq:SpinProj-0}) in $D$-dimensions the Breitenlohner-Maison-t'Hooft-Veltman (BMHV) scheme for treatment of $\gamma_5$ (see e.g.~\cite{Belusca-Maito:2020ala} for detailed discussion) had been used in the present calculation, relying on it's implementation in \texttt{FeynCalc}. Since $\gamma_5$ appears only in the definition of external states and not in the UV-divergent loops, one expects that the dependence on the scheme for treatment of $\gamma_5$ should cancel after the cancellation of IR divergences, if the same scheme is used both for virtual and real-emission corrections. So far no violation of this expectation had been reported in the literature, despite numerous NLO computations being done in the NRQCD-factorisation framework. 

Projecting on the CS or CO state of the $Q\bar{Q}$ boils down to the contraction of colour indices of heavy quarks with $\delta_{ij}/\sqrt{N_c}$ and $\sqrt{2}T_{ij}^a$ respectively. Finally, to match the definition of LDMEs~\cite{Bodwin:1994jh}, commonly used in NRQCD fits, the squared amplitude should be divided by the factor $2J+1$ and $2N_c$ for CS or $N_c^2-1$ for the CO case. All these conventional factors are included in expressions below. 

\begin{figure}
    \centering
    \includegraphics[width=0.7\textwidth]{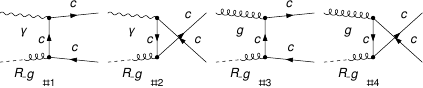}
    \includegraphics[width=0.35\textwidth]{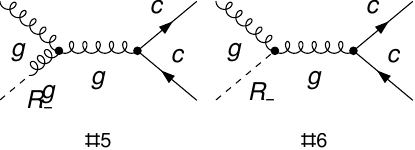}
    \caption{The LO diagrams contributing to impact-factors of processes (\ref{proc:ga+R->1S08}) -- (\ref{proc:g+R->3S18}). Dashed lines denote Reggeized gluons.}
    \label{fig:LO-diagrams}
\end{figure}

Tree-level Feynman diagrams contributing to the LO impact-factors of processes (\ref{proc:ga+R->1S08}) -- (\ref{proc:g+R->3S18}) are shown in the fig.~\ref{fig:LO-diagrams}. The diagrams \#1 and \#2 contribute to the process (\ref{proc:ga+R->1S08}). Diagrams \#3 and \#4 contribute to the processes (\ref{proc:g+R->1S01}) and (\ref{proc:g+R->1S08}). Due to the quantum numbers of the $Q\bar{Q}$-pair, the process (\ref{proc:g+R->3S18}) besides diagrams \#3 and \#4 also receives contributions from $s$-channel diagrams \#5 and \#6. The diagrams \#1 -- \#5 in the fig.~\ref{fig:LO-diagrams} contain only the $R_-g$ transition vertex (\ref{eq:Rg-vertex-EFT}), while the diagram \#6 contains the $R_-gg$ vertex (\ref{eq:Rgg-vertex-EFT}). 

The resulting LO amplitudes for the processes (\ref{proc:ga+R->1S08}) -- (\ref{proc:g+R->3S18}) are:
\begin{eqnarray}
    {\cal M}^{\text{(LO)}}_{\gamma R} \left[{}^1S_0^{[8]}\right] &=& \left( \frac{\left\langle {\cal O}\left[ {}^1S_0^{[8]} \right] \right\rangle}{M^3 (N_c^2-1)} \right)^{1/2}\frac{4i\sqrt{2} g_s e e_Q M \delta_{ab}}{M^2+{\bf k}_T^2} \epsilon^{\varepsilon(q) n_+ q p}, \label{eq:LO-ampl_ga+R->1S08} \\
    {\cal M}^{\text{(LO)}}_{g R} \left[{}^1S_0^{[1]}\right] &=& \left( \frac{\left\langle {\cal O}\left[ {}^1S_0^{[1]} \right] \right\rangle}{2M^3 N_c} \right)^{1/2}\frac{4i g_s^2 M \delta_{ac}}{\sqrt{N_c}(M^2+{\bf k}_T^2)} \epsilon^{\varepsilon(q) n_+ q p}, \label{eq:LO-ampl_g+R->1S01} \\
    {\cal M}^{\text{(LO)}}_{g R} \left[{}^1S_0^{[8]}\right] &=& \left( \frac{\left\langle {\cal O}\left[ {}^1S_0^{[8]} \right] \right\rangle}{M^3 (N_c^2-1)} \right)^{1/2}\frac{2i\sqrt{2} g_s^2 M d_{abc}}{M^2+{\bf k}_T^2} \epsilon^{\varepsilon(q) n_+ q p},  \label{eq:LO-ampl_g+R->1S08} \\
    {\cal M}^{\text{(LO)}}_{g R} \left[{}^3S_1^{[8]}\right] &=& \left( \frac{\left\langle {\cal O}\left[ {}^3S_1^{[8]} \right] \right\rangle}{3M^3 (N_c^2-1)} \right)^{1/2} \frac{i\sqrt{2}  g_s^2 f_{abc}}{M^2+{\bf k}_T^2} \varepsilon^*_\mu(p) \varepsilon_\nu(q)  \nonumber \\
    &&\times {\bf k}_T^2 \left[ 2q_+ g^{\mu\nu} + n_+^\nu \left( \frac{M^2+{\bf k}_T^2}{q_+} n_+^\mu - 2q^\mu \right) - 2n_+^\mu p^\nu \right],  \label{eq:LO-ampl_g+R->3S18}
\end{eqnarray}
where $a$ is the colour index of the $Q\bar{Q}^{[8]}$-pair and $b,c$ are colour indices of the gluon and Reggeon respectively. The obtained amplitudes vanish if one makes the replacement $\varepsilon(q)\to q$ for the gluon polarisation vector, i.e. they satisfy Slavnov-Taylor identity of QCD and the diagram \#6 in the fig.~\ref{fig:LO-diagrams} is necessary for this to be the case for the amplitude (\ref{eq:LO-ampl_g+R->3S18}). The results (\ref{eq:LO-ampl_g+R->1S01}) -- (\ref{eq:LO-ampl_g+R->3S18}) are related with the results for squared tree-level amplitudes of processes $R_+(q)+R_-(k)\to Q\bar{Q}\left[ {}^{2S+1}L_J^{[1,8]} \right]$ given in Ref.~\cite{Kniehl:2006sk} by the on-shell limit for the Reggeon $R_+(q)$: $q^2\to 0$.

\subsection{One-loop diagrams}

\begin{figure}
    \centering
    \parbox{0.5\textwidth}{\includegraphics[width=0.5\textwidth]{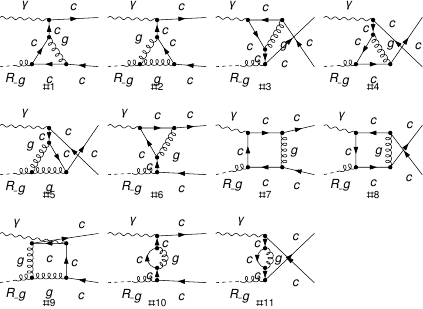}}\hspace{5mm}\parbox{0.4\textwidth}{\includegraphics[width=0.4\textwidth]{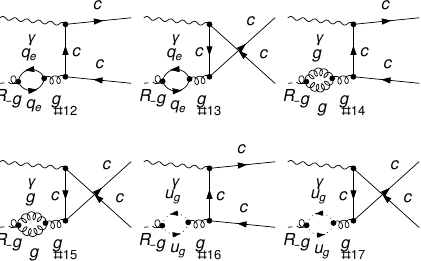}}
    \caption{One-loop diagrams with the $Rg$-coupling (\ref{eq:Rg-vertex-EFT}), contributing to the process (\ref{proc:ga+R->1S08}). Dashed lines denote Reggeised gluons, dotted lines -- Faddeev-Popov ghosts.}
    \label{fig:gamma+R->1S08:1L-Rg-diagrams}
\end{figure}

\begin{figure}
    \centering
    \includegraphics[width=0.4\textwidth]{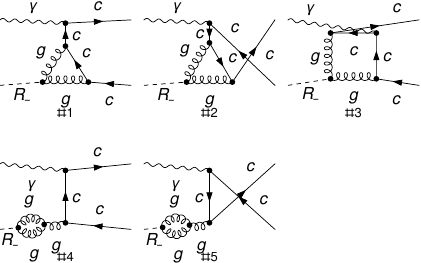}
    \caption{One-loop diagrams with the $Rgg$-coupling (\ref{eq:Rgg-vertex-EFT}), contributing to the process (\ref{proc:ga+R->1S08}).}
    \label{fig:gamma+R->1S08:1L-Rgg-diagrams}
\end{figure}

\begin{figure}
    \centering
    \includegraphics[width=0.7\textwidth]{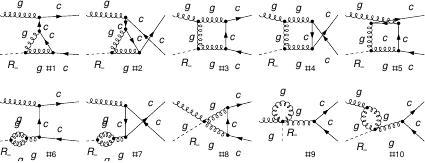}
    \includegraphics[width=0.7\textwidth]{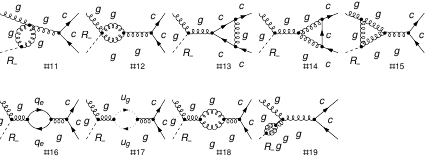}
    \caption{One-loop diagrams with the $Rgg$-coupling (\ref{eq:Rgg-vertex-EFT}), $Rggg$-coupling (\ref{Eq:R+_g3-vert}) and $Rgggg$-coupling (\ref{Eq:R+_g4-vert}), contributing to the processes (\ref{proc:g+R->1S01}),  (\ref{proc:g+R->1S08}) and  (\ref{proc:g+R->3S18}).}
    \label{fig:gR:1L-Rgg(g)-diagrams}
\end{figure}

At one loop, two kinds of diagrams contribute to the impact-factors. Those diagrams which contain only the $Rg$-coupling (\ref{eq:Rg-vertex-EFT}) are essentially the projections of ordinary one-loop QCD diagrams on a particular polarisation of an external gluon, and there is no technical issues with evaluating them. Diagrams from this class included into the impact-factor of the process (\ref{proc:ga+R->1S08}) are shown in the fig.~\ref{fig:gamma+R->1S08:1L-Rg-diagrams} as an example. All Feynman diagrams containing $Rg$-mixing vertex, except those which contain scaleless tadpoles or corrections to external {\it on-shell} quark and gluon legs should be taken into account. One also has to include the loop corrections to the Reggeon leg, such as diagrams \#12 -- \#14 in the fig.~\ref{fig:gamma+R->1S08:1L-Rg-diagrams}, following Refs.~\cite{Chachamis:2012cc,Chachamis:2012gh,Chachamis:2013hma,Nefedov:2019mrg}.  For the processes (\ref{proc:g+R->1S01}) -- (\ref{proc:g+R->3S18})  the diagrams with the $Rg$ vertex (\ref{eq:Rg-vertex-EFT}) had been taken into account following the same prescription, so they are not shown to save space. There are 38 such diagrams contributing to the process (\ref{proc:g+R->3S18}) at one loop, for example.

The second class of one-loop diagrams are those containing the induced vertices $Rgg$ (\ref{eq:Rgg-vertex-EFT}), $Rggg$ (\ref{Eq:R+_g3-vert}) and $Rgggg$ (\ref{Eq:R+_g4-vert}). For the process (\ref{proc:ga+R->1S08}) these diagrams are shown in the Fig.~\ref{fig:gamma+R->1S08:1L-Rgg-diagrams}. The diagrams \#4 and \#5 in the Fig.~\ref{fig:gamma+R->1S08:1L-Rgg-diagrams} are included as part of the correction to the external Reggeon leg, again following Refs.~\cite{Chachamis:2012cc,Chachamis:2012gh,Chachamis:2013hma,Nefedov:2019mrg}.  

Diagrams with $Rgg$ vertices \#1 -- \#7 in the Fig.~\ref{fig:gR:1L-Rgg(g)-diagrams} contribute to the processes (\ref{proc:g+R->1S01}) and (\ref{proc:g+R->1S08}), while all diagrams \#1 -- \#19 contribute to the process (\ref{proc:g+R->3S18}). Diagrams \#8, \#10, \#12 contain the $Rggg$ vertex (\ref{Eq:R+_g3-vert}) and the diagram \#9 contains the $Rgggg$ vertex (\ref{Eq:R+_g4-vert}). Despite its appearance, the diagram \#9 is not a scaleless tadpole, due to nontrivial structure of eikonal denominators in the vertex (\ref{Eq:R+_g4-vert}). Diagrams \#8, \#9, \#10, \#12 are important for the gauge-invariance of the impact-factor and for the consistency of the structure of rapidity divergences. The Mathematica package \texttt{FeynArts}~\cite{FeynArts}, equipped with the custom-made model-file, implementing vertices (\ref{eq:Rg-vertex-EFT}) -- (\ref{Eq:R+_g4-vert}) of the EFT, had been used to generate analytic expressions for above-described Feynman diagrams and various tools present in the \texttt{FeynCalc}~\cite{FeynCalc,Shtabovenko:2016sxi,Shtabovenko:2020gxv} framework had been employed to implement the tensor-reduction strategy explained in the next subsection.

\subsection{Tensor reduction}
\label{sec:reduction}
The following algorithm had been used to reduce the amplitudes, given by Feynman diagrams described above, down to scalar one-loop integrals. For each tensor integral of rank $n$ which is identified in the expression for the Feynman diagram (i.e. an integral containing the product $l_{\mu_1}\ldots l_{\mu_n}$ in the numerator), an ansatz for the tensor structure is written in terms of independent ``external'' four-vectors $k_\mu$, $p_\mu$ and $\tilde{n}_\mu^+$ and projectors on each of the basis elements in this ansatz are derived. Applying this projectors to the tensor integrals at hands, allows one to express the amplitude as a linear combination of integrals with scalar products of independent external four-vectors and the loop momentum $l$ in the numerator. These scalar products can be expressed in terms of the denominators of the loop integral and resulting set of integrals with various positive and negative powers of denominators can be reduced to a small set of master integrals by the standard integration-by-parts (IBP) reduction~\cite{Chetyrkin:1981qh} software, the \texttt{FIRE}~\cite{Smirnov:2019qkx} had been used for this purpose in the present calculation through the convenient interface to \texttt{FeynCalc} provided by \texttt{FeynHelpers}~\cite{Shtabovenko:2016whf}.   

However there is a well-known obstacle for the application of IBP reduction, which arises when dealing with certain one-loop diagrams in quarkonium physics. Due to the prescription $p_{Q}=p_{\bar{Q}}=p/2$, some denominators of the Feynman integral become linearly-dependent. Linear dependence arises e.g. for the diagram \#8 and \#9 in the Fig.~\ref{fig:gamma+R->1S08:1L-Rg-diagrams}  and \#3 in the Fig.~\ref{fig:gamma+R->1S08:1L-Rgg-diagrams} as well as e.g. for the diagrams \#1 and \#13 in the Fig.~\ref{fig:gR:1L-Rgg(g)-diagrams}. The linear relation between denominators ($D_i$) takes the form:
\begin{equation}
    \sum\limits_{i=1}^{n_{\text{lin. dep. den.}}} c_i D_i = a, \label{eq:lin-rel-Di}
\end{equation}
with $n_{\text{lin. dep. den.}}$ being less or equal to the total number of propagators in the loop and coefficients $c_i$ and $a$ being independent from the loop momentum $l$. If, as it happens e.g. in the case of diagram \#9 in the Fig.~\ref{fig:gamma+R->1S08:1L-Rg-diagrams}, the parameter $a\neq 0$, then one can divide the Eq.~(\ref{eq:lin-rel-Di}) by $a$ to put it into a form of the decomposition of unity. If $a=0$, which happens for the so-called {\it Coulomb-divergent} diagrams, where the final-state heavy-quark lines are connected by a gluon, e.g. diagram \#8 in the Fig.~\ref{fig:gamma+R->1S08:1L-Rg-diagrams}, then one have to divide the Eq.~(\ref{eq:lin-rel-Di}) by one of it's nonzero terms $c_i D_i$ to obtain the decomposition of unity. Either way, by putting this unity into the numerator of the corresponding diagram and expanding the expression,  one obtains the sum of terms in each of which the linear dependence of denominators had been resolved. The linear dependence of eikonal denominators, arising e.g. in the diagram \#9 in the Fig.~\ref{fig:gR:1L-Rgg(g)-diagrams} is resolved in the same way.  For the case of $a=0$, this {\it partial-fractioning} procedure leads to squaring of one of the denominators, but the resulting scalar integrals are again reduced to master integrals with denominators with only first powers by the IBP software.  

 If the relative momentum of $Q$ and $\bar{Q}$ had been kept nonzero in the present computation, than the Coulomb divergence in the diagrams where outgoing heavy quark lines are connected by a gluon, would manifest itself as the pole $\sim \alpha_s C_F/(4\pi v)$ in the relative velocity of heavy quarks $v$. But since the $v=0$ limit had been taken from the very beginning in the present calculation, the Coulomb divergence is put to zero by dimensional regularisation just as any other power-law divergence. The remaining finite part of the amplitude corresponds to the properly defined NLO in $\alpha_s$ Wilson coefficient in the NRQCD matching formula, as explained e.g. in Sec. 6 of Ref.~\cite{Petrelli:1997ge}. 

All the manipulations  described above as well as IBP-reduction, has to be done retaining at least $O(r)$-terms in the numerators, since one encounters power-divergent scalar integrals $\propto r^{-1+\epsilon}$ as well as Gram determinants $\propto r$. Terms $O(r^2)$ and higher can be systematically dropped at each step, reducing the size of expressions.

As a result of the procedure, described above, the amplitude is reduced to a linear combination of a small number of scalar one-loop integrals with or without linear denominators. The scalar integrals with linear denominators but without masses in quadratic propagators, which are needed for the present calculation, are the same as listed in Ref.~\cite{Nefedov:2019mrg}. The new element of the present computation are the rapidity-divergent integrals with internal massive line, and the next section is devoted to their computation.

\section{Computation of rapidity-divergent master integrals with masses}
\label{sec:master-ints}
\subsection{Partial fractioning of mass dependence in presence of linear denominator}

In the present computation one encounters scalar integrals with one linear denominator and one, two or three quadratic denominators. The following general notation for them is used in the present paper:
\begin{eqnarray}
    A_{[+]}(k_1,m_1^2)&=&\int\frac{[d^d l]}{[\tilde{l}_+] ((l+k_1)^2-m_1^2+i\eta)}, \\
    B_{[+]}(k_1,m_1^2;k_2,m_2^2)&=&\int\frac{[d^d l]}{[\tilde{l}_+] ((l+k_1)^2-m_1^2+i\eta) ((l+k_2)^2-m_2^2+i\eta)}, \label{eq:B[+]-def}\\
    C_{[+]}(k_1,m_1^2;k_2,m_2^2)&=&\int\frac{[d^d l]}{[\tilde{l}_+] (l^2+i\eta) ((l+k_1)^2-m_1^2+i\eta) ((l+k_2)^2-m_2^2+i\eta)}, \label{eq:C[+]-def}
\end{eqnarray}

where $d=4-2\epsilon$ where the integration measure
\begin{eqnarray}
&&[d^d l]=\frac{(\mu^2)^\epsilon d^d l}{i\pi^{d/2} r_{\Gamma}},  \\
&&r_\Gamma = \frac{\Gamma^2(1-\epsilon) \Gamma(1+\epsilon)}{\Gamma(1-2\epsilon)} = \frac{1}{\Gamma(1-\epsilon)}+O(\epsilon^3).
\end{eqnarray}

The results for such master integrals with massless internal lines and up to two scales of virtuality are collected in Ref.~\cite{Nefedov:2019mrg}. Integrals with one scale of virtuality, given in~\cite{Nefedov:2019mrg} had been computed for the first time in Refs.~\cite{Hentschinski:2011tz,Chachamis:2012cc}. However, in heavy-quarkonium production calculations, which are the subject of the present paper, one encounters for the first time the following rapidity-divergent scalar integrals with up to two massive internal lines:
\begin{eqnarray}
&&    A_{[+]}\left(\frac{p}{2},\frac{M^2}{4}\right), \\
&&    B_{[+]}\left( p, 0; \frac{p}{2}, \frac{M^2}{4} \right),\ \ B_{[+]}\left( k,0; \frac{p}{2}, \frac{M^2}{4} \right), \ \ B_{[+]}\left( 0,0; \frac{p}{2}-k, \frac{M^2}{4} \right),  \label{eq:ints_B+_1m} \\
&&  B_{[+]}\left( \frac{p}{2},\frac{M^2}{4}; k-\frac{p}{2}, \frac{M^2}{4} \right), \label{eq:int_B+_2m} \\
&& C_{[+]}\left( q,0 ; \frac{p}{2}-k, \frac{M^2}{4} \right), \ \  C_{[+]}\left( k,0 ; \frac{p}{2}, \frac{M^2}{4} \right), \label{eq:ints_C+_1m} \\
&& C_{[+]}\left( \frac{p}{2},\frac{M^2}{4} ; k-\frac{p}{2}, \frac{M^2}{4} \right). \label{eq:int_C+_2m}
\end{eqnarray}

In the remainder of this subsection the strategy used to compute these integrals is described. If one considers the eikonal and massive denominators together, disregarding other quadratic denominators in the integral, then the following algebraic identity is true:
\begin{eqnarray}
 && \frac{1}{[\tilde{n}_+ \cdot l + X_+] (l^2-m^2+i\eta)} = \frac{1}{[\tilde{n}_+\cdot l + X_+] ((l+\rho \tilde{n}_+)^2+i\eta)} \nonumber \\
     && + 2\rho \frac{\tilde{n}_+\cdot l + \frac{m^2}{2\rho} + \frac{\rho \tilde{n}_+^2}{2}}{ [\tilde{n}_+\cdot l + X_+] ((l+\rho \tilde{n}_+)^2+i\eta) (l^2-m^2+i\eta)},  \label{eq:mass-id}
\end{eqnarray}
where the massive denominator can always be put in a form $l^2-m^2$ by the shift of the loop momentum $l$, which just leads to a change of the constant $X_+$.  The key observation is, that the parameter $\rho$ in Eq.~(\ref{eq:mass-id}) can be chosen to solve the equation $m^2/\rho + \rho \tilde{n}_+^2 = 2X_+$, which obviously leads to the cancellation of the linear denominator in the second term. Since $\tilde{n}_+^2=4r \ll 1$, one should better choose the solution of the quadratic equation for $\rho$ which is not singular in the limit $r\to 0$ to avoid dealing with extra singularities in $r$. As a result, in the first term of Eq.~(\ref{eq:mass-id}) the massive denominator is traded for a massless one, while in the second term both remaining denominators are quadratic. If more than one massive denominator is present, like in the integral (\ref{eq:int_B+_2m}) or (\ref{eq:int_C+_2m}), then the identity (\ref{eq:mass-id}) can be applied recursively, until elimination of all massive denominators from the integral with the linear propagator is achieved. As a result of this procedure, any integral with linear denominator and several massive propagators can be expressed as linear combination of integrals with only quadratic propagators and one integral with the eikonal denominator, all quadratic propagators in which are massless.  Having achieved that, one can expand integrals with quadratic propagators in the $r\ll 1$ limit, using known results for them, while the remaining integral with linear propagator might still have to be computed. The interface of the \texttt{Package-X}~\cite{Patel:2015tea} to \texttt{FeynHelpers} is used in the present calculation to obtain analytic results for standard scalar integrals with quadratic denominators.

As an example of this procedure, let's consider one of integrals (\ref{eq:ints_B+_1m}), which can be written as:
\begin{eqnarray*}
    B_{[+]}\left(p,0; \frac{p}{2}, \frac{M^2}{4} \right) = \int \frac{[d^d l]}{ [\tilde{n}_+\cdot l - \tilde{p}_+/2] (l^2 - M^2/4) (l+p/2)^2 }, 
\end{eqnarray*}
choosing $\rho = \left(-\tilde{p}_+ + \sqrt{\tilde{p}_+^2 - 4M^2 r}\right)/(8r) = -M^2/(4p_+) +O(r)$ in the relation (\ref{eq:mass-id}), one gets:
\begin{eqnarray}
   B_{[+]}\left(p,0; \frac{p}{2}, \frac{M^2}{4} \right) &=& B_{[+]}\left(p,0; \frac{P}{2}, 0 \right) \nonumber \\ 
   && + \frac{M^2}{2p_+} \int\frac{[d^D l]}{ (l+\rho \tilde{n}_+)^2 \left(l^2-\frac{M^2}{4}\right) \left( l + \frac{p}{2} \right)^2 } + O(r), \label{eq:mass-id-example}
\end{eqnarray}
where the new vector $P=p+2\rho \tilde{n}_+=p-M^2 n_+/(2p_+)+O(r)$ is exactly light-like ($P^2=0$) for any $r$. The loop integral with quadratic denominators in Eq.~(\ref{eq:mass-id-example}) corresponds to the standard Passarino-Veltman function: 
\begin{eqnarray*}
   && C\left( (\rho \tilde{n}_+)^2, \left(\frac{p}{2}\right)^2,\left(\rho \tilde{n}_+ - \frac{p}{2}\right)^2;0,\frac{M^2}{4},0 \right) \\ 
   && =  C\left(\frac{M^4r}{4p_+^2},\frac{M^2}{4},\frac{M^2}{2};0,\frac{M^2}{4},0\right)+O(r), 
\end{eqnarray*}
where the $O(r)$ virtuality of the first leg of the triangle had been retained, because the expansion in $r\ll 1$ is done before the expansion in $\epsilon\ll 1$. This generates some terms $O(\ln r)$ and $O(\ln^2 r)$ which need to be taken into account. Now, all what is left to do is to compute the new rapidity-divergent integral with massless denominators -- $B_{[+]}\left(p,0; {P}/{2}, 0\right)$, which have appeared in the r.h.s. of the Eq.~(\ref{eq:mass-id-example}). 


It turns out, that after above-described procedure, all rapidity-divergent scalar integrals which are needed for the computation are either listed in Ref.~\cite{Nefedov:2019mrg} or can be expressed in terms of new integrals which are computed in the Sec.~\ref{sec:new-master-B} and~\ref{sec:new-master-C} of the present paper.

\subsection{Master integral with two quadratic propagators\label{sec:new-master-B}}

In the process of reduction, as described above the integral $B_{[+]}(k_1,0;k_2,0)$ arises, in the notation of Eq.~(\ref{eq:B[+]-def}). For integrals with linear denominators the parametric representation with all parameters ranging from zero to infinity turns-out to be more convenient:
\[
\frac{1}{A^{n_1} B^{n_2}} = \frac{\Gamma(n_1+n_2)}{\Gamma(n_1) \Gamma(n_2)} \int\limits_0^{\infty} \frac{x^{n_2-1} dx }{(A+Bx)^{n_1+n_2}},
\]
which leads to:
\begin{eqnarray}
    && B_{(+)}(\tilde{k}_1^+,Y,\xi,r) = \int \frac{[d^d l]}{(\tilde{l}^++i\eta) ((l+k_1)^2 + i\eta) ((l+k_2)^2+i\eta) } \nonumber \\
    && = -\frac{\Gamma(1-2\epsilon)}{\Gamma^2 (1-\epsilon)}  \left(\frac{\mu}{\tilde{k}_1^+} \right)^{2\epsilon} \int\limits_0^\infty \frac{dx_1 dx_2}{\tilde{k}_1^+} (x_1+x_2)^{-1+2\epsilon} [x_1+\xi x_2 + Yx_1x_2 + r]^{-1-\epsilon}, \label{eq:B(+)-def}
\end{eqnarray}
where $Y=-(k_1-k_2)^2/(\tilde{k}_1^+)^2-i\eta$ and $\xi = \tilde{k}_2^+/\tilde{k}_1^+$. In the present paper the case $\xi\neq 0$ is treated, while for the case of $\xi=0$ this integral is given in the Sec.~3.2 of Ref.~\cite{Nefedov:2019mrg}. To obtain the result with the PV prescription for the eikonal pole, one uses the following analytic continuation formula:
\begin{equation}
    B_{[+]}(\tilde{k}_1^+,Y,\xi,r) = \frac{1}{2}\left[ B_{(+)}(\tilde{k}_1^+-i\eta,Y,\xi-i\eta,r) + B_{(+)}(\tilde{k}_1^+-i\eta,Y,\xi+i\eta,r) \right].
\end{equation}

The integral (\ref{eq:B(+)-def}) can be expanded in $r\ll 1$ using Mellin-Barnes representation (see e.g.~\cite{Smirnov}). The exact-in-$\epsilon$ result is:
\begin{eqnarray}
    B_{(+)}(\tilde{k}_1^+,Y,\xi,r)&=&\frac{1}{\tilde{k}_1^+} \left(\frac{\mu}{\tilde{k}_1^+} \right)^{2\epsilon} \left[ \frac{Y^{-\epsilon}}{\xi \epsilon^2} {}_2F_1 \left(1,1-\epsilon,1+\epsilon;\frac{1}{\xi} \right)  - \frac{\sec(\pi\epsilon) r^{\epsilon}}{2\epsilon^2} \frac{1-\xi^{-2\epsilon}}{1-\xi} \right. \nonumber \\
    &&\left.- \frac{(4Y)^{-\epsilon} \Gamma(1+\epsilon) \Gamma\left(\frac{1}{2} - \epsilon \right)}{\epsilon^2 \sqrt{\pi} } \xi^{-\epsilon} (\xi-1)^{-1-2\epsilon} \right]+O(r^{-\epsilon}, r),
\end{eqnarray}
where the terms which scale as $r^{-\epsilon}$ had been put to zero, assuming $\epsilon<0$. The $\epsilon$-expansion of the above integral can be done with the help of the \texttt{HypExp} package~\cite{Huber:2005yg}, the answer is:
\begin{equation}
    B_{(+)}(\tilde{k}_1^+,Y,\xi,r) = \frac{1}{2\tilde{k}_1^+ (1-\xi)} \left[ 4{\rm Li}_2\left(\frac{\xi-1}{\xi} \right) + \ln\xi \left( 3\ln \xi -2\ln r - 2\ln Y \right) \right] + O(r^{-\epsilon},r,\epsilon). 
\end{equation}

\subsection{Master integral with three quadratic propagators\label{sec:new-master-C}}

Certain integrals which arise in the process of reduction, turn out to be particular cases of the integral $C_{[+]}(k_1,0;k_2,0)$ in the notation of Eq.~(\ref{eq:C[+]-def}) with $k_1^2=0$. The parametric representation for the integral of this kind, can be written as:
\begin{eqnarray}
    &&C_{(+)}(\tilde{k}_1^+,\kappa,\xi,r,T)= \int \frac{[d^d l]}{(\tilde{l}_+ + i\eta) (l^2+i\eta)((l+k_1)^2+i\eta) ((l+k_2)^2+i\eta)} \label{eq:C(+)-def} \\
    &&= \frac{\Gamma(1-2\epsilon)\Gamma(2+\epsilon)}{\Gamma(1+\epsilon)\Gamma^2(1-\epsilon)} \left(\frac{\mu^2}{-k_2^2} \right)^{\epsilon} \int\limits_0^\infty \frac{dx_1 dx_2 dx_3}{(-k_2^2)^{3/2}}(x_1+x_2+x_3)^{2\epsilon} \nonumber \\
    &&\times \left[ \kappa (x_2+\xi x_3) +x_3(x_1+ Tx_2) + r \right]^{-2-\epsilon},
\end{eqnarray}
where $T=[(k_1-k_2)^2+i\eta]/[k_2^2+i\eta]$, $\kappa=\tilde{k}_1^+/\sqrt{-k_2^2}$, $\xi=\tilde{k}_2^+/\tilde{k}_1^+$ and for $\xi \neq 0$ one has the relation:
\begin{equation}
C_{(+)}(\tilde{k}_1^+,\kappa,\xi,r,T) = \frac{1}{\kappa\xi} C_{(+)}(\tilde{k}_1^+,\xi^{-1},\xi,r\kappa^{-2}\xi^{-2},T), \label{eq:C(+)-rel}    
\end{equation}
which follows from the rescaling of $x_{1,2,3}\to \kappa\xi  x_{1,2,3}$.

The integral with PV pole prescription is obtained with the help of the following analytic continuation formula:
\begin{equation}
   C_{[+]}(\tilde{k}_1^+,\kappa,\xi,r,T)=\frac{1}{2}\left[ C_{(+)}(\tilde{k}_1^+,\kappa-i\eta,\xi-i\eta,r,T) - C_{(+)}(\tilde{k}_1^+,-\kappa-i\eta,\xi-i\eta,r,T) \right].
\end{equation}
One also notices that for the case of $\xi=0$ the integral is equivalent to the integral in Eq. (34) in the Sec. 3.2 of the Ref.~\cite{Nefedov:2019mrg} so only $\xi\neq 0$ case is discussed in the present paper.

This integral depends on more parameters than the previous one, so treating it with multiple Mellin-Barnes representation becomes less convenient. Instead a version of the method of differential equations~\cite{Kotikov:1990kg} delivers the result. Differentiating the integral (\ref{eq:C(+)-def}) w.r.t. variable $T$ and IBP-reducing the result down to the same set of MIs using \texttt{FIRE}~\cite{Smirnov:2019qkx}, one obtains an in-homogeneous differential equation of the form:
\begin{eqnarray}
    \left[ \frac{d}{dT} - \frac{\epsilon \xi}{(1-\xi) + T\xi} \right] C_{[+]}(\tilde{k}_1^+,\kappa,\xi,r,T) = J(T), \label{eq:C(+)-DE}
\end{eqnarray}
where the r.h.s. of the Eq.~(\ref{eq:C(+)-DE}) is expressed in terms of master integrals as:
\begin{eqnarray}
   && J(T)= \frac{(1-2\epsilon) r}{(-k_2^2) \kappa^2 T} \left[ \frac{1}{\xi}A_{[+]}(k_1,0) - \frac{1-\xi}{(1-\xi)+\xi T} A_{[+]}(k_2,0) \right] \label{eq:DE-C[+]-rhs} \\ 
   && -\frac{\epsilon }{(-k_2^2) T [(1-\xi) + \xi T]} \left[ \xi^2 B_{[+]}(k_2,0;0,0) + (1-\xi^2) B_{[+]}(k_1,0;k_2,0) \right] \nonumber \\
   && -\frac{(1-2\epsilon) }{(-k_2^2) \tilde{k}_1^+ }\frac{(2-\xi + \xi T) B(k_2^2,0) - (1+T - \xi (1-T)) B((k_1-k_2)^2,0)} {T (1-T) [(1-\xi) + \xi T]} + O(r), \nonumber
\end{eqnarray}
where one have to retain the contributions of integrals $A_{[+]}$ because they are proportional to $r^{-1+\epsilon}$ (Ref.~\cite{Nefedov:2019mrg}, Eq. (26) in Sec. 3.1) while all other $O(r)$-suppressed contributions can be dropped. The standard notation $B$ is used for the one-loop functions $B(k^2,m^2)=\int [d^dl]/[l^2((l+k)^2-m^2)]$ in the Eq.~(\ref{eq:DE-C[+]-rhs}).

The in-homogeneous equation (\ref{eq:C(+)-DE}) has the following general solution:
\begin{equation}
    C_{[+]}(\tilde{k}_1^+,\kappa,\xi,r,T) = \left(1+\frac{\xi T}{1-\xi} \right)^{\epsilon} C_{[+]}(\tilde{k}_1^+,\kappa,\xi,r,0) + \int\limits_0^{T} dt \left( \frac{(1-\xi) + \xi T}{(1-\xi) + \xi t} \right)^{\epsilon} J(t), \label{eq:C[+]-DE-sol}
\end{equation}
moreover, substituting the results for master integrals one finds that the terms $O(r^\epsilon)$ cancel between the first and the second lines of Eq.~(\ref{eq:DE-C[+]-rhs}), leading to the following $r$-independent result for $J(T)$:
\begin{eqnarray}
 && \left(\frac{\mu^2}{-k_2^2}\right)^{-\epsilon}  (-k_2^2)^{3/2} J(T) = \frac{(1-\xi) 4^{-\epsilon } \xi ^{-\epsilon -1} T^{-1-\epsilon}}{ \kappa  (1-T)  [(1-\xi)+ T\xi ]} \left[ \frac{\xi  (\xi -1)^{2 \epsilon }}{\sqrt{\pi}\epsilon } \Gamma \left(\frac{1}{2}-\epsilon \right) \Gamma (1+\epsilon
   ) \right. \nonumber \\
   && \left. - \frac{4^{\epsilon } \xi ^{\epsilon }}{\epsilon} \left((\xi -1) \, _2F_1\left(1,1-\epsilon
   ;\epsilon +1;\frac{1}{\xi }\right)-\xi \right) \right] \nonumber \\
   && + \frac{4^{-\epsilon } \xi ^{-\epsilon -1} T ^{-\epsilon }}{ \kappa  (1-T) [(1-\xi)+ T\xi ]} \left[ \frac{\xi(\xi -1)^{2 \epsilon
   +1}}{\sqrt{\pi} \epsilon } \Gamma \left(\frac{1}{2}-\epsilon \right) \Gamma (1+\epsilon ) \right. \nonumber \\
   && \left. -\frac{4^{\epsilon } \xi^{\epsilon }}{\epsilon} \left((\xi -1)^2 \,
   _2F_1\left(1,1-\epsilon ;\epsilon +1;\frac{1}{\xi }\right)-\xi  \left(\xi -2 T ^{\epsilon }+1\right)\right) \right].
\end{eqnarray}
Another important feature of this result is that all hypergeometric functions in it do not depend on $T$, only on $\xi$, therefore the $T$-dependence of $J(T)$ is entirely given in terms of elementary functions.

Substituting thus obtained $J(T)$ to the solution (\ref{eq:C[+]-DE-sol}) one obtains a convenient representation for the $T$-dependent integral, which can be straightforwardly expanded in $\epsilon$ using the formula:
\[
\int\limits_0^T dt\ t^{-1-\epsilon} f(t) = -\frac{T^{-\epsilon}}{\epsilon} f(0) +   \int\limits_0^T dt\ t^{-1-\epsilon} [f(t)-f(0)],
\]
in such a way one completely recovers the $T$-dependence of the integral in question. 

All what remains to do is to find the initial condition at $T=0$ for the equation (\ref{eq:C(+)-DE}), which can be done using two-fold Mellin-Barnes representation for the integral (\ref{eq:C(+)-def}) at $T=0$. The answer is:
\begin{eqnarray}
    && C_{(+)} (\tilde{k}_1^+,\kappa,\xi,r,0) = \left(\frac{\mu^2}{-k_2^2} \right)^{\epsilon} \frac{1}{(-k_2^2) \tilde{k}_1^+ \xi } \frac{2\pi \csc(\pi \epsilon)}{\Gamma(1-\epsilon)} \nonumber \\ && \times \left[ \frac{1}{\Gamma(2+\epsilon)} {}_2F_1 \left( 1,1;2+\epsilon; \frac{1}{\xi} \right) +  \xi (\xi-1)^\epsilon \frac{\Gamma(1-\epsilon)}{\epsilon} \right] + O(r^{-\epsilon},r), 
\end{eqnarray}
where the $O(r^{-\epsilon})$ terms again have been neglected. In Ref.~\cite{Nefedov:2019mrg} such terms are retained and it is shown that they cancel in the final expression for the amplitude, however one also can rely on the argument that $r^{-\epsilon}\to 0$ for $\epsilon<0$ and therefore it is sufficient to check the cancellation of $r^{\epsilon}$-dependence in the amplitude, systematically dropping $O(r^{-\epsilon})$-terms from all master integrals.

Substituting the initial condition thus obtained to the solution (\ref{eq:C[+]-DE-sol}) and expanding it in $\epsilon$ one obtains the following remarkably simple result for the $T$-dependent integral:
\begin{eqnarray}
    && C_{[+]}(\tilde{k}_1^+,\kappa,\xi,r,T) = \frac{1}{(-k_2^2) \tilde{k}_1^+} \left( \frac{\mu^2}{-k_2^2-i\eta} \right)^{\epsilon} \left[ \frac{1}{\epsilon^2} + \frac{ \ln [(T-i\eta)(\xi-i\eta)] }{\epsilon} \right. \nonumber \\
    && \hspace{-10mm} \left. -2 {\rm Li}_2(1-T+i\eta) +\frac{1}{2} \ln^2 [(\xi-i\eta) (T-i\eta)] - \ln^2 (T-i\eta) + \frac{\pi^2}{3}   \right]+O(r^{-\epsilon},r,\epsilon)
\end{eqnarray}
Trivial dependence of this result on $\kappa$ is explained by the relation (\ref{eq:C(+)-rel}).

\section{Results and cross-checks}
\label{sec:results}

\subsection{$\gamma R\to Q\bar{Q}[{}^1S_0^{[8]}]$}
The Lorentz structure of the one-loop amplitude for the process (\ref{proc:ga+R->1S08}) coincides with the Lorentz structure of the LO amplitude (\ref{eq:LO-ampl_ga+R->1S08}). Since the quark-mass renormalisation counterterm is proportional to the derivative of the LO amplitude (\ref{eq:LO-ampl_ga+R->1S08}) with respect to the quark mass in the internal propagator in the diagrams of Fig.~\ref{fig:LO-diagrams}:
\begin{equation}
    \frac{\partial^{\text{(int.)}}}{\partial m_Q}  {\cal M}^{\text{(LO)}}_{\gamma R} \left[{}^1S_0^{[8]}\right] = -\frac{2M}{M^2+{\bf k}_T^2}  {\cal M}^{\text{(LO)}}_{\gamma R} \left[{}^1S_0^{[8]}\right], \label{eq:1S0-mQ-CT}
\end{equation}
it is convenient to add to all the results presented below the counterterm $\delta Z_{m_Q}\times \partial^{\rm (int.)}_{m_Q} {\cal M}^{\text{(LO)}}$ in the on-shell scheme at one loop, i.e. with:
\begin{equation}
    \delta Z_{m_Q}= -\frac{\bar{\alpha}_s}{4\pi}\left[\frac{3C_F}{\epsilon}+2C_F\left( 2 + \frac{3}{2} \ln \frac{4\mu^2}{M^2} \right)\right], \label{eq:dZ_mQ-OS}
\end{equation}
where $\bar{\alpha}_s=g_s^2 r_\Gamma (\mu_R^2)^{-\epsilon}/(4\pi)^{1-\epsilon}$. After that, the divergence structure of the result becomes proportional just to the LO amplitude. The on-shell scheme of Eq.~(\ref{eq:dZ_mQ-OS}) is a de-facto standard renormalisation scheme in heavy-quarkonium production physics. Wave-function renormalisation and $\alpha_s$-counterterms have {\it not} been added to the results presented below. Moreover, the internal closed quark loops, such as diagrams \# 12 and \#13 in the Fig.~\ref{fig:gamma+R->1S08:1L-Rg-diagrams}, take into account only $n_F$ {\it massless} flavours.

After substituting master integrals, significant simplifications are possible, using well-known identities for dilogarithm functions which allow one to express the amplitude in terms of just two dilogarithms: ${\rm Li}_2(-\tau)$ and ${\rm Li}_2(-1-2\tau)$. In the present paper only the real part of the coefficients is given, which is all what is needed for the calculation of cross sections in the NLL approximation. The divergent and finite parts of the coefficient in front of the LO Lorentz structure in the one-loop amplitude of the process (\ref{proc:ga+R->1S08}) can be separated as:
\begin{eqnarray}
    &&\text{Re}\,C[\gamma R\to {}^1S_0^{[8]}]= \frac{\bar{\alpha}_s}{4\pi}\Biggl\{ \left( \frac{\mu^2}{{\bf k}_T^2}\right)^\epsilon  \frac{1}{\epsilon}\left( C_A\ln \frac{q_+^2}{r M^2} + \beta_0 + 3C_F - 2C_A \right)  \nonumber \\
    && -\frac{10}{9} n_F + C_F  C[\gamma R\to {}^1S_0^{[8]},C_F] +  C_A C[\gamma R\to {}^1S_0^{[8]},C_A] \Biggr\}+O(r,\epsilon), \label{eq:gamma+R-1S08:CLO-expansion}
\end{eqnarray}
where $\tau={\bf k}_T^2/M^2$ and $\beta_0=11C_A/3-2n_F/3$. The coefficient in front of $C_F$ is\footnote{These expressions have been put into an explicitly real-valued form for $\tau>0$ and supersede the expressions published in Ref.~\cite{Nefedov:2023uen}. Real part of those expressions may be ambiguous due to analytic continuation of logarithms and dilogarithms which is not properly specified in Ref.~\cite{Nefedov:2023uen}.}:
\begin{eqnarray}
   && C[\gamma R\to {}^1S_0^{[8]},C_F]= \frac{1}{6 (\tau +1)^2} \Biggl\{ -12 \tau  (\tau
   +1) \text{Li}_2(-2 \tau -1) + \frac{6 L_2}{\tau } (-2 L_1 \tau +L_1+6 \tau  (\tau +1)) \nonumber \\
   && +\frac{1}{(2 \tau +1)^2} \Biggl[ (\tau +1) 12
   \ln(2) (\tau +1) \left(6 \tau ^2+8 \tau +3\right)-8 \tau ^3 \left(9 \ln (\tau +1)+2 \pi
   ^2+15\right) \nonumber \\
   && -4 \tau ^2 \left(30 \ln (\tau +1)+\pi ^2+63\right)+8 \tau  \left(-6 \ln (\tau +1)+\pi
   ^2-21\right) \nonumber \\
   && +18 (\tau +1) (2 \tau +1)^2 \ln (\tau )+3 \pi ^2-36 \Biggr]\Biggr\},\label{eq:gamma+R-1S08:CF-coef}
\end{eqnarray}
where $L_1=L_1^{(+)}-L_1^{(-)}-L_2/2$ with $L_{1}^{(\pm)}=\sqrt{\tau(1+\tau)}\ln\left( \sqrt{1+\tau} \pm \sqrt{\tau} \right)$ and $L_2=\sqrt{\tau(1+\tau)}\ln\left( 1+2\tau+2\sqrt{\tau(1+\tau)} \right)$.

In the coefficient in front of $C_A$ it is convenient to separate the part containing dilogarithms from the rest of the expression:
\begin{equation}
    C[\gamma R\to {}^1S_0^{[8]},C_A]= C^{({\rm Li}_2)}[\gamma R\to {}^1S_0^{[8]},C_A]+ C^{(\ln)}[\gamma R\to {}^1S_0^{[8]},C_A].\label{eq:gamma+R-1S08:CA-coef}
\end{equation}

The part containing dilogarithms has the form:
\begin{eqnarray}
   && C^{({\rm Li}_2)}[\gamma R\to {}^1S_0^{[8]},C_A] =\frac{2 (\tau  (\tau  ((\tau -4) \tau -6)-4)+1)}{(\tau -1) (\tau
   +1)^3}\text{Li}_2(-\tau ) \nonumber \\
   && -\frac{(\tau -1) \left(\tau ^2+\tau +1\right)}{(\tau +1)^3} \text{Li}_2(-2 \tau -1),
\end{eqnarray}
while the remainder is:
\begin{eqnarray}
    && C^{(\ln)}[\gamma R\to {}^1S_0^{[8]},C_A] = \frac{L_2 (L_2 (2 \tau -1)-12 \tau  (\tau +1))}{4 \tau  (\tau +1)^2} \nonumber \\
    && +\frac{1}{36 (\tau -1) (\tau +1)^3 (2 \tau +1)} \Biggl\{ 8 \left(\tau ^2-1\right) (9 \ln(2) \tau  (\tau  (\tau +5)+2)+38 \tau  (\tau  (2
   \tau +5)+4)+38) \nonumber \\
   && +18 (2 \tau +1) \ln (\tau ) [-2 \tau^4+(\tau  ((\tau -4) \tau -6)-4)
   \tau  \ln (\tau )+\ln (\tau )+2] \nonumber \\
   &&+72 \tau  \left(\tau  \left(3 (\tau +2) \tau
   ^2+\tau -5\right)-4\right) \ln (\tau +1)+3 \pi ^2 (2 \tau +1) (\tau  ((\tau -1) \tau  (16
   \tau +21)-25)+1) \nonumber \\
   && -72 \ln (\tau +1) \Biggr\}.
\end{eqnarray}

The asymptotics of these expressions for $\tau\ll 1$ is useful for the study of the issues of (in-)stability of NLO corrections in HEF~\cite{Nefedov:2020ecb}:
\begin{eqnarray}
&& C[\gamma R\to {}^1S_0^{[8]},C_F] =   3 \ln (\tau )+\frac{\pi ^2}{2}-6+6 \ln (2) + O(\tau), \\
&& C[\gamma R\to {}^1S_0^{[8]},C_A]  =-\frac{1}{2} \log ^2(\tau )-\log (\tau )-\frac{\pi ^2}{6}+\frac{76}{9}+O(\tau).
\end{eqnarray}

\begin{figure}
    \centering
    \includegraphics[width=0.49\linewidth]{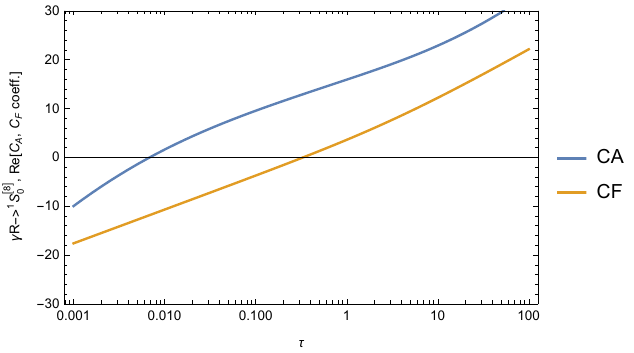}\includegraphics[width=0.49\linewidth]{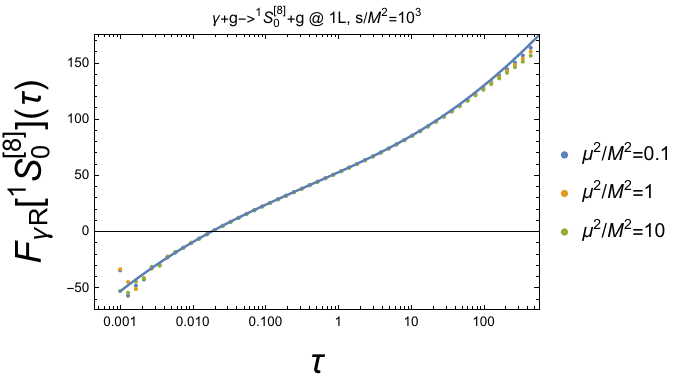}
    \caption{Left panel: plots of the coefficients (\ref{eq:gamma+R-1S08:CF-coef}) and (\ref{eq:gamma+R-1S08:CA-coef}) as functions of $\tau={\bf k}_T^2/M^2$. Right panel: plots of the quantity $F_{\gamma R}[{}^1S_0^{[8]}]=C_F C[\gamma R\to {}^1S_0^{[8]},C_F]+C_A C[\gamma R\to {}^1S_0^{[8]},C_A]$ as function of $\tau$. Points in the right panel are obtained from the exact one-loop amplitude of the process (\ref{test-proc:gamma+g-1S08}) for $s=10^3 M^2$ as described in the main text. Solid line -- Eqns. (\ref{eq:gamma+R-1S08:CF-coef})+(\ref{eq:gamma+R-1S08:CA-coef})}
    \label{fig:gamma+g-1S08_plots}
\end{figure}

Having obtained the result (\ref{eq:gamma+R-1S08:CLO-expansion}) in the EFT, it is important to discuss how it is related with the high-energy limit of the one-loop amplitudes in QCD. For example, if one considers the process:
\begin{equation}
    \gamma(q) + g_{\lambda_1}(k_1) \to Q\bar{Q}\left[ {}^1S_0^{[8]}\right](p) + g_{\lambda_2}(k_2), \label{test-proc:gamma+g-1S08}
\end{equation}
at one loop, then the Regge limit $s=(k_1+q)^2 \gg -t=-(q-p)^2$ of the real part of the contribution with antisymmetric colour-octet (${\bf 8}_a$) exchange in the $t$-channel\footnote{Which is the only possible colour structure for the process (\ref{test-proc:gamma+g-1S08}) but becomes important for processes (\ref{test-proc:g+g-1S08}) and (\ref{test-proc:g+g-3S18}). Taking the interference with tree-level amplitude automatically projects-out needed colour and helicity structures in all cases considered in the present paper.} to the amplitude with $\lambda_1=\lambda_2=\pm$, independently on helicities of other particles, is proportional to the Regge limit of the corresponding tree-level amplitude and can be decomposed in terms of the EFT ingredients as follows~\cite{Hentschinski:2011tz,Chachamis:2012cc,Nefedov:2019mrg}:
\begin{equation}
\frac{\text{Re }{\cal M}^{\text{(1-loop)}}_{\text{proc. (\ref{test-proc:gamma+g-1S08})}}}{{\cal M}^{\text{(Born)}}_{\text{proc. (\ref{test-proc:gamma+g-1S08})}}} = \text{Re}\,C[\gamma R\to {}^1S_0^{[8]}] + \gamma_{gRg}^{(1)}(\sqrt{s},-t) - \Pi^{(1)}(-t) + O(-t/s,M^2/s), \label{eq:QCD-EFT:gamma+g-1S08+g}
\end{equation}
where $\gamma_{gRg}^{(1)}(\sqrt{s},-t)$ is the one-loop correction to the $g+R\to g$-scattering vertex in the EFT~\cite{Chachamis:2012cc,Nefedov:2019mws}, which is closely related to the one-loop correction to the impact-factor of a gluon~\cite{Fadin:1992zt,Fadin:1993wh,DelDuca:1998kx,Fadin:1999de}, and $\Pi^{(1)}$ is the one-loop correction to the Reggeon propagator in the EFT~\cite{Chachamis:2012cc}. The required combination of these quantities reads:
\begin{eqnarray}
&&  \hspace{-5mm}  \gamma_{gRg}^{(1)}(\sqrt{s},-t) - \Pi^{(1)}(-t) = \frac{\bar{\alpha}_s}{4\pi} \Biggl\{ -\frac{2C_A}{\epsilon^2} + \frac{1}{\epsilon} \left[ C_A \ln \frac{s r}{-t} - 2C_A \ln \frac{\mu^2}{-t} -\frac{8}{3}C_A + \frac{2}{3} n_F \right] \nonumber \\ 
&& \hspace{-5mm}+\frac{1}{9} \left[-3 \ln \left(-\frac{t}{\mu ^2}\right) \left(C_A (6 \ln (r)-5)+2 n_F\right)-31 C_A+10 n_F\right] \nonumber \\
&& \hspace{-5mm}+\frac{1}{2} C_A
   \left[-2 \ln \left(-\frac{\mu ^2}{t}\right) \left(\ln (r)-\ln \left(-\frac{s}{t}\right)+\ln \left(-\frac{\mu ^2}{t}\right)+1\right)+\pi
   ^2-4\right] + O(r,\epsilon)\Biggr\}. \label{eq:gRg-vertex}
\end{eqnarray}
Note that the $\ln r$-divergence cancels in the Eq.~(\ref{eq:QCD-EFT:gamma+g-1S08+g}) and the quantity (\ref{eq:gRg-vertex}) is factorised (i.e. independent) from the other scattering vertex ($C[\ldots+R\to\ldots]$) in the process under consideration. The last statement means that the Eq.~(\ref{eq:QCD-EFT:gamma+g-1S08+g}) holds also for the Regge limits of one-loop amplitudes of test processes (\ref{test-proc:g+g-1S01}), (\ref{test-proc:g+g-1S08}) and (\ref{test-proc:g+g-3S18}) below, up to replacement of the corresponding one-loop corrections to the quarkonium production vertices but with the quantity (\ref{eq:gRg-vertex}) being unchanged.

The numerical check of the equality (\ref{eq:QCD-EFT:gamma+g-1S08+g}) for the finite parts of the amplitude is presented in the right panel of the Fig.~\ref{fig:gamma+g-1S08_plots}. The finite part of the interference between the one-loop corrected and tree-level {\it exact} amplitudes of the process (\ref{test-proc:gamma+g-1S08}) 
had been computed numerically for $s=10^3 M^2$ at several values of $\tau=-t/M^2$ and $\mu^2$ using \texttt{FormCalc}~\cite{Hahn:2000jm}, as described in Appendix~\ref{appendix:FormCalc}. From these numerical data, the quantity $F_{\gamma R}[{}^1S_0^{[8]}]=C_F C[\gamma R\to {}^1S_0^{[8]},C_F]+C_A C[\gamma R\to {}^1S_0^{[8]},C_A]$ had been deduced with the help of Eqns.~(\ref{eq:QCD-EFT:gamma+g-1S08+g}), (\ref{eq:gRg-vertex})  and (\ref{eq:gamma+R-1S08:CLO-expansion})  and plotted with points in the right panel of the Fig.~\ref{fig:gamma+g-1S08_plots}, while the analytic result for this quantity, given by Eqns.~(\ref{eq:gamma+R-1S08:CF-coef}) and (\ref{eq:gamma+R-1S08:CA-coef}) is plotted with the solid line. Note, that trivial logarithms of $\tau$ which follow from the structure of divergent part, as well as the $\mu$-dependence, is removed from the plotted quantity due to it's definition, Eq.~(\ref{eq:gamma+R-1S08:CLO-expansion}).  As one can see from the right panel of the Fig.~\ref{fig:gamma+g-1S08_plots}, at very low values of $\tau<0.004$ numerical points are scattered due to instability of the \texttt{FormCalc} computation. At $\tau\sim s/M^2$ one expects the power corrections $O(-t/s)$ to kick in, which explains why the points deviate from the analytic expectation (solid line) at large $\tau$ and the $\mu$-independence of the plotted quantity is violated\footnote{I.e. points of different colours in the right panel of the Fig.~\ref{fig:gamma+g-1S08_plots}, corresponding to different $\mu^2$ in the numerical calculation, do not lie on the same curve for these very large values of $\tau$.}. In a wide range of moderate values of $\tau$ there is an agreement between the numerical data and the analytic result up to power-corrections $O(M^2/s)$ and $O(-t/s)$. One should note, that calculation of the exact $2\to 2$ amplitude of the process (\ref{test-proc:gamma+g-1S08}) does not rely on IBP-reduction, as explained in Appendix~\ref{appendix:FormCalc}, which turns the obtained agreement into a very strong check, supporting the correctness of the EFT calculation.

\subsection{$gR\to Q\bar{Q}[{}^1S_0^{[1]}]$}

As for the previous case, the only Lorentz structure one obtains for the one-loop amplitude of the process (\ref{proc:g+R->1S01}) is the LO one, Eq.~(\ref{eq:LO-ampl_g+R->1S01}). The quark mass renormalisation counterterm is related with the LO amplitude in the same way as in Eq.~(\ref{eq:1S0-mQ-CT}), multiplied by $\delta Z_{m_Q}$, and it have already been added to the results presented below. After substituting master integrals, the real part of the coefficient in front of the LO Lorentz structure (\ref{eq:LO-ampl_g+R->1S01}) can be written as follows:
\begin{eqnarray}
    &&\text{Re}\,C[gR\to {}^1S_0^{[1]}]= \frac{\bar{\alpha}_s}{4\pi}\Biggl\{ \left( \frac{\mu^2}{{\bf k}_T^2}\right)^\epsilon \left[-\frac{C_A}{\epsilon^2}+ \frac{1}{\epsilon}\left( C_A\ln \frac{q_+^2}{r {\bf k}_T^2} + \beta_0 + 3C_F - C_A \right) \right] \nonumber \\
    && -\frac{10}{9} n_F + C_F  C[gR\to {}^1S_0^{[1]},C_F] +  C_A C[gR\to {}^1S_0^{[1]},C_A] \Biggr\}+O(r,\epsilon), \label{eq:g+R-1S01:CLO-expansion}
\end{eqnarray}

For the process (\ref{proc:g+R->1S01}), as it was already noted in Ref.~\cite{Nefedov:2023uen}, the coefficient in front of $C_F$ coincides with the same coefficient for the one-loop amplitude of the process (\ref{proc:ga+R->1S08}):
\begin{equation}
    C[gR\to {}^1S_0^{[1]},C_F]=C[\gamma R\to {}^1S_0^{[8]},C_F],
\end{equation}
which is given by Eq.~(\ref{eq:gamma+R-1S08:CF-coef}).

The coefficient in front of $C_A$ again can be split into parts containing ${\rm Li}_2$ and the remainder:
\begin{equation}
    C[g R\to {}^1S_0^{[1]},C_A]= C^{({\rm Li}_2)}[g R\to {}^1S_0^{[1]},C_A]+ C^{(\ln)}[g R\to {}^1S_0^{[1]},C_A],\label{eq:g+R-1S01:CA-coef}
\end{equation}
where the part containing dilogarithms is:
\begin{eqnarray}
&& C^{({\rm Li}_2)}[g R\to {}^1S_0^{[1]},C_A]=\frac{2 (\tau  (\tau  (\tau  (7 \tau +8)+2)-4)-1)}{(\tau -1) (\tau +1)^3}\text{Li}_2(-\tau ) \nonumber \\
&& -\frac{ \tau  (\tau  (4 \tau +5)+3)}{(\tau +1)^3}\text{Li}_2(-2 \tau -1),
\end{eqnarray}
while the remainder, containing logarithms and rational functions, is:
\begin{eqnarray}
&& C^{(\ln)}[g R\to {}^1S_0^{[1]},C_A]=-\frac{L_2^2}{2 \tau  (\tau +1)^2} \nonumber \\
&& + \frac{1}{18 (\tau -1) (\tau +1)^3} \Biggl\{-2 \left(\tau ^2-1\right) (18 \ln(2) (\tau -1) \tau -67 (\tau +2) \tau -67) \nonumber \\
&&+18 [\ln (\tau ) \left(-2 \tau ^4+\left(\tau  \left(-\tau ^3+\tau
   +3\right)+2\right) \tau  \ln (\tau )+2 \tau ^2+\ln (\tau )\right) \nonumber \\
&& -(\tau -1)^2 (\tau +1)^3 \ln ^2(\tau +1)+2 (\tau -1) (\tau +1)^2 \left(\tau +(\tau +1)^2 \ln (\tau
   )\right) \ln (\tau +1)] \nonumber \\
&&+\pi ^2 (3 \tau  (\tau  (\tau  (15 \tau +14)-3)-12)-6) \Biggr\},
\end{eqnarray}
with $L_2$ being defined after Eq.~(\ref{eq:gamma+R-1S08:CF-coef}). At $\tau\ll 1$ the coefficient in front of $C_A$ behaves as:
\begin{equation}
   C[g R\to {}^1S_0^{[1]},C_A]= -\ln ^2(\tau )+\frac{\pi ^2}{3}+\frac{67}{9} + O(\tau).
\end{equation}

\begin{figure}
    \centering
    \includegraphics[width=0.49\linewidth]{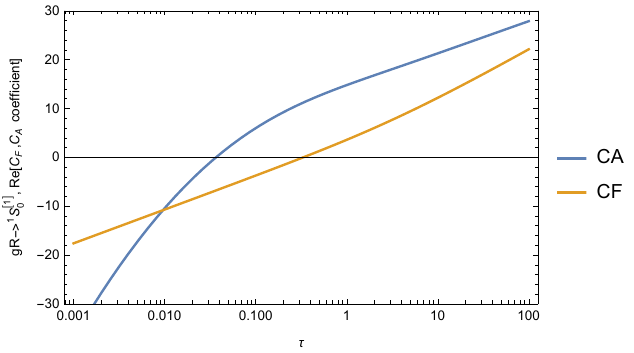}\includegraphics[width=0.49\linewidth]{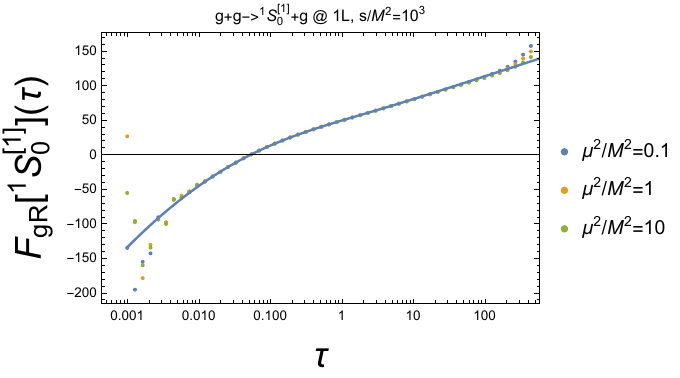}
    \caption{Left panel: plots of the coefficients (\ref{eq:gamma+R-1S08:CF-coef}) and (\ref{eq:g+R-1S01:CA-coef}) as functions of $\tau={\bf k}_T^2/M^2$. Right panel: plots of the quantity $F_{gR}[{}^1S_0^{[1]}]=C_F C[gR\to {}^1S_0^{[1]},C_F]+C_A C[gR\to {}^1S_0^{[1]},C_A]$ as function of $\tau$. Points in the right panel are obtained from the exact one-loop amplitude of the process (\ref{test-proc:g+g-1S01}) for $s=10^3 M^2$ as described in the main text. Solid line -- Eqns. (\ref{eq:gamma+R-1S08:CF-coef})+(\ref{eq:g+R-1S01:CA-coef})}
    \label{fig:g+g-1S01_plots}
\end{figure}

Plots of the $C_F$ and $C_A$ coefficients are shown in the left panel of the Fig.~\ref{fig:g+g-1S01_plots}, while on the right panel, the numerical results for the $F_{gR}[{}^1S_0^{[1]}]=C_F C[gR\to {}^1S_0^{[1]},C_F]+C_A C[gR\to {}^1S_0^{[1]},C_A]$, obtained from the Regge limit of the exact one-loop amplitude of the process:
\begin{equation}
    g(q) + g(k_1) \to Q\bar{Q}\left[ {}^1S_0^{[1]}\right](p) + g(k_2), \label{test-proc:g+g-1S01}
\end{equation}
using Eqns.~(\ref{eq:QCD-EFT:gamma+g-1S08+g}) and (\ref{eq:g+R-1S01:CLO-expansion}) are plotted with points, together with analytic result, given by Eqns.~(\ref{eq:gamma+R-1S08:CF-coef}) and (\ref{eq:g+R-1S01:CA-coef}), shown by the solid line. The interpretation of these results is the same as for the right panel of the Fig.~\ref{fig:gamma+g-1S08_plots}: at $\tau<0.005$ the numerical data disagree with the analytic result due to the numerical instability, while at $\tau \sim s/M^2$ power corrections $O(-t/s)$ spoil the agreement and $\mu$-independence. At moderate values of $\tau$, numerical data agree with the analytic curve with expected accuracy, confirming the results (\ref{eq:g+R-1S01:CA-coef}) and (\ref{eq:gamma+R-1S08:CF-coef}).  

\subsection{$gR\to Q\bar{Q}[{}^1S_0^{[8]}]$}

The one-loop amplitude for the process (\ref{proc:g+R->1S08}) is once again proportional to the LO result. The $m_Q$-renormalisation counterterm had been added to it, as was done also for the process (\ref{proc:ga+R->1S08}) and (\ref{proc:g+R->1S01}), such that the divergent part of the result is just proportional to the LO amplitude. The one-loop coefficient takes the form:
\begin{eqnarray}
    &&\text{Re}\,C[gR\to {}^1S_0^{[8]}]= \frac{\bar{\alpha}_s}{4\pi}\Biggl\{ \left( \frac{\mu^2}{{\bf k}_T^2}\right)^\epsilon \left[-\frac{C_A}{\epsilon^2}+ \frac{1}{\epsilon}\left( C_A\ln \frac{q_+^2}{r {\bf k}_T^2} + C_A\ln \left(1+\frac{{\bf k}_T^2}{M^2} \right) \right.\right.  \nonumber \\
    && + \beta_0 + 3C_F - 2C_A \biggr) \biggr] -\frac{10}{9} n_F  \nonumber \\
    && + C_F  C[gR\to {}^1S_0^{[8]},C_F] +  C_A C[gR\to {}^1S_0^{[8]},C_A] \Biggr\}+O(r,\epsilon). \label{eq:g+R-1S08:CLO-expansion}
\end{eqnarray}

The piece of the finite part, proportional to $C_F$, once again turns out to be the same as for the case of the process (\ref{proc:ga+R->1S08}):
\begin{equation}
C[gR\to {}^1S_0^{[8]},C_F]=C[\gamma R\to {}^1S_0^{[8]},C_F],
\end{equation}
given by Eq.~(\ref{eq:gamma+R-1S08:CF-coef}), while the term proportional to $C_A$ can be split into di-logarithmic part and the remainder:
\begin{equation}
    C[g R\to {}^1S_0^{[8]},C_A]= C^{({\rm Li}_2)}[g R\to {}^1S_0^{[8]},C_A]+ C^{(\ln)}[g R\to {}^1S_0^{[8]},C_A],\label{eq:g+R-1S08:CA-coef}
\end{equation}
with the former one being equal to:
\begin{eqnarray}
&& C^{({\rm Li}_2)}[g R\to {}^1S_0^{[8]},C_A]=\frac{4 \tau  (\tau  (2 \tau +3)+2)}{(\tau +1)^3}\text{Li}_2(-\tau) \nonumber \\
&&-\frac{\tau  (\tau  (3 \tau +4)+3)}{(\tau +1)^3}\text{Li}_2(-2 \tau -1),
\end{eqnarray}
and the remainder piece being equal to:
\begin{eqnarray}
 && C^{(\ln)}[g R\to {}^1S_0^{[8]},C_A]=\frac{L_2 (L_1^{(+)} (8 \tau -4) +(1-2 \tau)L_2-12 \tau  (\tau +1))}{4 \tau  (\tau +1)^2} \nonumber \\
 &&+ \frac{1}{36 (\tau +1)^3} \Biggl\{ 4 (\tau +1) (36 \ln(2) \tau +85 (\tau +2) \tau +85)-18 [\tau  (\tau +1) \left(\tau ^2+3\right)+2] \ln ^2(\tau ) \nonumber \\
 && +36 \ln (\tau
   ) [(\tau +1)^3 (\tau +2) \ln (\tau +1)-(\tau +1) \left(2 \tau ^2+\tau +1\right)] +3 \pi ^2 (\tau  (\tau  (24 \tau +41)+26)+1) \nonumber \\
   &&+18 (\tau
   +1)^2 \ln (\tau +1) \left(8 \tau -(\tau +1)^2 \ln (\tau +1)+4\right) \Biggr\}.
\end{eqnarray}

The $\tau\ll 1$ asymptotics of the $C_A$-coefficient is:
\begin{equation}
  C[g R\to {}^1S_0^{[8]},C_A]=  -\ln ^2(\tau )-\ln (\tau )+\frac{\pi ^2}{12}+\frac{85}{9}+ O(\tau).
\end{equation}

\begin{figure}
    \centering
    \includegraphics[width=0.49\linewidth]{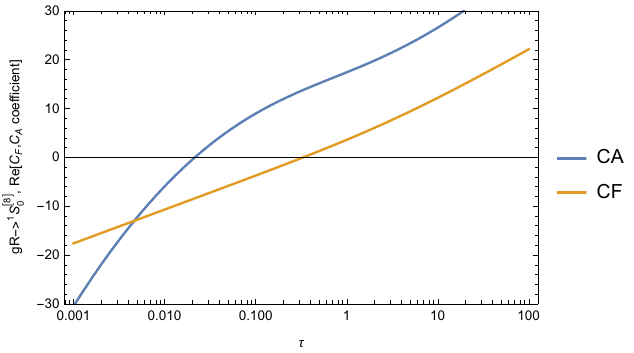}\includegraphics[width=0.49\linewidth]{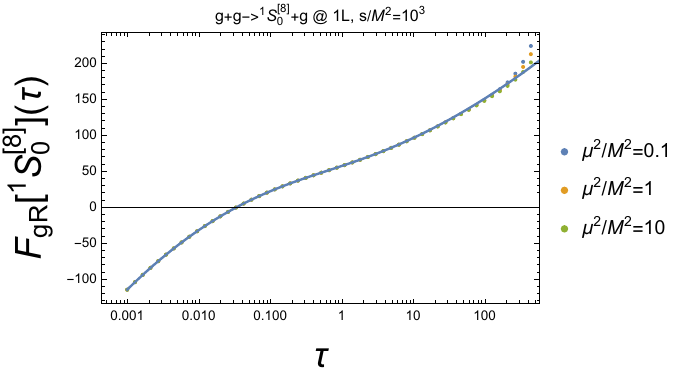}
    \caption{Left panel: plots of the coefficients (\ref{eq:gamma+R-1S08:CF-coef}) and (\ref{eq:g+R-1S08:CA-coef}) as functions of $\tau={\bf k}_T^2/M^2$. Right panel: plots of the quantity $F_{gR}[{}^1S_0^{[8]}]=C_F C[gR\to {}^1S_0^{[8]},C_F]+C_A C[gR\to {}^1S_0^{[8]},C_A]$ as function of $\tau$. Points in the right panel are obtained from the exact one-loop amplitude of the process (\ref{test-proc:g+g-1S08}) for $s=10^3 M^2$ as described in the main text. Solid line -- Eqns. (\ref{eq:gamma+R-1S08:CF-coef})+(\ref{eq:g+R-1S08:CA-coef})}
    \label{fig:g+g-1S08_plots}
\end{figure}

Plots of the Eqns.~(\ref{eq:g+R-1S08:CA-coef}) and (\ref{eq:gamma+R-1S08:CF-coef}) are shown in the left panel of Fig.~\ref{fig:g+g-1S08_plots}. It is worth noting that both $C_A$-coefficients for the production of ${}^1S_0^{[1]}$ and ${}^1S_0^{[8]}$ states are similar in their numerical magnitude. As have already been done for other cases, the numerical test of obtained results is performed by extracting the quantity $F_{gR}[{}^1S_0^{[8]}]=C_F  C[gR\to {}^1S_0^{[8]},C_F] +  C_A C[gR\to {}^1S_0^{[8]},C_A]$ from numerical results for the interference of the one-loop corrected and tree-level amplitudes of the process:
\begin{equation}
    g(q) + g(k_1) \to Q\bar{Q}\left[ {}^1S_0^{[8]}\right](p) + g(k_2), \label{test-proc:g+g-1S08}
\end{equation}
with the help of Eqns.~(\ref{eq:QCD-EFT:gamma+g-1S08+g}) and (\ref{eq:g+R-1S08:CLO-expansion}). Comparison of these numerical results to the corresponding analytic expectations is shown in the right panel of the Fig.~\ref{fig:g+g-1S08_plots} and one again observes a good agreement between numerical results and the prediction of the Eqns.~(\ref{eq:g+R-1S08:CA-coef}) and (\ref{eq:gamma+R-1S08:CF-coef}) for all values of $\tau=-t/M^2$, except for the very large ones $\tau\sim s/M^2 = 10^3$, where the $O(-t/s)$ power corrections to the full one-loop QCD amplitude spoil the agreement. The numerical instability at very low $\tau$ is much less pronounced this time.

\subsection{$gR\to Q\bar{Q}[{}^3S_1^{[8]}]$}
The $O(\alpha_s)$ amplitude for the process (\ref{proc:g+R->3S18}) has more complicated structure than for the previous cases. It can be expressed as a linear combination of Lorentz structures which includes the LO amplitude (\ref{eq:LO-ampl_g+R->3S18}), it's derivative with respect to the quark mass in the internal propagator\footnote{This Lorentz structure violates the Ward identity w.r.t. incoming on-shell gluon and the gauge-invariance of the amplitude will be restored only after the addition of the mass counterterm.}:
\begin{eqnarray}
    \frac{\partial^{\text{(int.)}}}{\partial m_Q}  {\cal M}^{\text{(LO)}}_{g R} \left[{}^3S_1^{[8]}\right] &=& \left( \frac{\left\langle {\cal O}\left[ {}^3S_1^{[8]} \right] \right\rangle}{3M^3 (N_c^2-1)} \right)^{1/2} \frac{2\sqrt{2}i g_s^2 f_{abc}}{(M^2+{\bf k}_T^2)^2} \varepsilon^*_\mu(p) \varepsilon_\nu(q)  \nonumber \\  
    &\times& [(M^2-{\bf k}_T^2) ( q_+ g^{\mu\nu} -  n_+^\mu p^\nu) - 2M^2n_+^\nu q^\mu] , \label{eq:Mass-der:gR-3S18}
\end{eqnarray}
and the following two gauge-invariant Lorentz structures, arising at one loop:
\begin{eqnarray}
    &&{\cal M}^{\text{(LS-1)}}_{g R} \left[{}^3S_1^{[8]}\right] =   \left( \frac{\left\langle {\cal O}\left[ {}^3S_1^{[8]} \right] \right\rangle}{3M^3 (N_c^2-1)} \right)^{1/2} \frac{i\sqrt{2} g_s^2 f_{abc}}{M^2+{\bf k}_T^2} \varepsilon^*_\mu(p) \varepsilon_\nu(q) \nonumber \\
    &&\times \Bigl[2q_+ {\bf k}_T^2 g^{\mu \nu} + q^\mu\Bigl( n_+^\nu ((d-2)M^2+(d-4){\bf k}_T^2) - 2(d-2) q_+ p^\nu \Bigr) \nonumber \\
    && + \frac{{\bf k}_T^2 n_+^\mu}{q_+} \Bigl( (M^2+{\bf k}_T^2)n_+^\nu - 2q_+p^\nu \Bigr) \Bigr], \label{eq:gR-3S18:LS1}\\
    &&{\cal M}^{\text{(LS-2)}}_{g R} \left[{}^3S_1^{[8]}\right] =   \left( \frac{\left\langle {\cal O}\left[ {}^3S_1^{[8]} \right] \right\rangle}{3M^3 (N_c^2-1)} \right)^{1/2} \frac{i\sqrt{2} g_s^2 f_{abc}}{M^2+{\bf k}_T^2} \varepsilon^*_\mu(p) \varepsilon_\nu(q) \nonumber \\
    &&\times \Bigl[2q_+ {\bf k}_T^2 g^{\mu \nu} + q^\mu\Bigl( n_+^\nu ((d-2)M^2+(d-4){\bf k}_T^2) - 2(d-2) q_+ p^\nu \Bigr)\Bigr], \label{eq:gR-3S18:LS2}
\end{eqnarray}
which are chosen to be orthogonal to the LO amplitude (\ref{eq:LO-ampl_g+R->3S18}) and to each-other, i.e. their interference with the LO amplitude or with each-other is equal to zero. Appearance of these new Lorentz structures is akin to the appearance of the Lorentz-structure violating the helicity conservation in the gluon impact-factor at NLO~\cite{Fadin:1992zt,Fadin:1993wh,DelDuca:1998kx,Fadin:1999de,Chachamis:2012cc,Nefedov:2019mws}. The scalar coefficients in front of these Lorentz structures can be expanded in terms of master integrals.

After substituting master integrals into the scalar coefficients, significant simplifications are possible, using dilogarithm identities. The divergent and finite parts of the coefficient in front of the LO Lorentz structure for the ${}^3S_1^{[8]}$ production can be separated as:
\begin{eqnarray}
    &&{\rm Re}\,C[{}^3S_1^{[8]},\text{LO}]= \frac{\bar{\alpha}_s}{4\pi}\Biggl\{ \left( \frac{\mu^2}{{\bf k}_T^2}\right)^\epsilon \left[ -\frac{C_A}{\epsilon^2} + \frac{1}{\epsilon}\left( C_A\ln \frac{q_+^2 (1+\tau)}{r{\bf k}_T^2} + \beta_0 + 3C_F - 2C_A  \right) \right] \nonumber \\ 
    && + n_F  C[{}^3S_1^{[8]},\text{LO},n_F] + C_F  C[{}^3S_1^{[8]},\text{LO},C_F] +  C_A C[{}^3S_1^{[8]},\text{LO},C_A] \Biggr\}+O(r,\epsilon). \label{eq:g+R-3S18:CLO-expansion}
\end{eqnarray}
While, as one would expect since these Lorentz structures do not contribute to the Born order, the coefficients in front of the Lorentz structures~(\ref{eq:gR-3S18:LS1}) and~(\ref{eq:gR-3S18:LS2}) are free from IR, UV and rapidity divergences and one can decompose them according to the colour structures as:
\begin{eqnarray}
    &&C[{}^3S_1^{[8]},\text{LS-1}]= n_F  C[{}^3S_1^{[8]},\text{LS-1},n_F] + C_F  C[{}^3S_1^{[8]},\text{LS-1},C_F] \nonumber \\
    && +  C_A C[{}^3S_1^{[8]},\text{LS-1},C_A]+O(r,\epsilon), \\
    &&C[{}^3S_1^{[8]},\text{LS-2}]= n_F  C[{}^3S_1^{[8]},\text{LS-2},n_F] + C_F  C[{}^3S_1^{[8]},\text{LS-2},C_F] \nonumber \\ 
    && +  C_A C[{}^3S_1^{[8]},\text{LS-2},C_A]+O(r,\epsilon).
\end{eqnarray}

For the LO Lorentz structure, the coefficient in front of $C_F$ is:
\begin{eqnarray}
  && C[{}^3S_1^{[8]},\text{LO},C_F] =\frac{2 (\tau  (\tau +6)-1) \text{Li}_2(-2 \tau -1)}{(\tau +1)^3} + \frac{L_2^2 (\tau  (\tau +5)-2)}{\tau  (\tau +1)^4}-\frac{4 L_2 (\tau +7)}{(\tau +1)^3}  \nonumber \\
  && +\frac{2 \ln(2) (\tau  (\tau  (2 \tau  (6 \tau
   +19)+83)+56)+11)}{\left(2 \tau ^2+3 \tau +1\right)^2} +\frac{2}{2 \tau +1}+3 \ln (\tau )-4 \nonumber \\
  && +\frac{4 (\tau  (\tau  (\tau +22)+19)+4) \ln
   (\tau +1)}{\left(2 \tau ^2+3 \tau +1\right)^2}+\frac{\pi ^2 (\tau +5) (2 \tau -1)}{3 (\tau +1)^3}, \label{eq:3S18:LO-CF}
\end{eqnarray}
and the corresponding coefficient in front of $n_F$ is:
\begin{equation}
    C[{}^3S_1^{[8]},\text{LO},n_F] = -\frac{
    2 \left(5 \tau ^2+13 \tau +8\right)+\left(6 \tau ^2+9 \tau +9\right) \ln (\tau )}{9 (\tau +1)^2}.
\end{equation}
The $C_A$ coefficient of the LO Lorentz structure can be decomposed into a part containing dilogarithms and a remainder:
\begin{eqnarray}
   {\rm Re}\, C[{}^3S_1^{[8]},\text{LO},C_A] &=& C^{(\text{Li}_2)}[{}^3S_1^{[8]},\text{LO},C_A] + C^{(\ln)}[{}^3S_1^{[8]},\text{LO},C_A] 
\end{eqnarray}
where
\begin{eqnarray}
&& C^{(\text{Li}_2)}[{}^3S_1^{[8]},\text{LO},C_A] = \frac{2 (\tau -1) \left(2 \tau ^3+3 \tau ^2+6 \tau +1\right)}{(\tau +1)^5}\text{Li}_2(-\tau ) \nonumber \\
&&-\frac{3 \tau ^4+9 \tau ^3+15 \tau ^2-\tau -2}{(\tau +1)^5}  \text{Li}_2(-2
   \tau -1), \label{eq:3S18:LO-CA-Li2}
\end{eqnarray}
and
\begin{eqnarray}
&&C^{(\ln)}[{}^3S_1^{[8]},\text{LO},C_A] = \frac{L_2 (-4 L^{(+)}_1 \tau  (\tau +5)+8 L^{(+)}_1+L_2 \tau  (\tau +5)-2 L_2+4 \tau  (\tau +1) (\tau +7))}{2 \tau  (\tau +1)^4} \nonumber \\
&&+\frac{1}{18 (\tau +1)^5 (2 \tau +1)}\Biggl\{ 12 \ln(2) (\tau  (\tau  (\tau  (4 \tau  (4 \tau +13)+59)-25)-27)-3) (\tau +1) \nonumber \\
&&+(2 \tau +1) \left(3 \pi ^2 [(\tau  (2 \tau  (\tau
   +6)+9)+11) \tau ^2+\tau +5]+2 (\tau  (73 \tau +140)+55) (\tau +1)^3\right) \nonumber \\
&& -9 (2 \tau +1) (\tau  (\tau  (\tau +1) (\tau  (\tau +2)+7)+10)+2)
   \ln ^2(\tau ) \nonumber \\
&& +72 (\tau  ((\tau -4) \tau -4)-1) (\tau +1)^2 \ln (\tau +1) \nonumber \\
&& +6 (2 \tau +1) (\tau +1) \ln (\tau ) [\tau  (\tau  (\tau  (8 \tau
   +19)+29)-3)+3 (\tau +1)^4 \ln (\tau +1)+3] \Biggr\}.\label{eq:3S18:LO-CA-ln}
\end{eqnarray}

For the Lorentz structures~(\ref{eq:gR-3S18:LS1}) and~(\ref{eq:gR-3S18:LS2}) the coefficients are real-valued and again can be expanded in terms of colour factors $C_A$, $C_F$ and $n_F$, which in turn can be separated into the part containing dilogarithms and the remainders:
\begin{eqnarray}
    C[{}^3S_1^{[8]},\text{LS-1/2},C_A/C_F/n_F] &=& C^{(\text{Li}_2)}[{}^3S_1^{[8]},\text{LS-1/2},C_A/C_F/n_F] \nonumber \\ + C^{(\ln)}[{}^3S_1^{[8]},\text{LS-1/2},C_A/C_F/n_F],
\end{eqnarray}
where:
\begin{eqnarray}
    &&\hspace{-1cm}C^{(\text{Li}_2)}[{}^3S_1^{[8]},\text{LS-1},C_F]=-\frac{2 \left(\tau ^2+6 \tau -1\right) \text{Li}_2(-2 \tau -1)}{(\tau +1)^3}=- C^{(\text{Li}_2)}[{}^3S_1^{[8]},\text{LO},C_F],  \label{eq:3S18:LS1-CF-Li2} \\
    &&\hspace{-1cm}C^{(\text{Li}_2)}[{}^3S_1^{[8]},\text{LS-1},C_A]=- C^{(\text{Li}_2)}[{}^3S_1^{[8]},\text{LO},C_A], \label{eq:3S18:LS1-CA-Li2} \\
   &&\hspace{-1cm}C^{(\text{Li}_2)}[{}^3S_1^{[8]},\text{LS-2},C_F]= \frac{2 (2 \tau -1) \text{Li}_2(-2 \tau -1)}{(\tau +1)^2}, \label{eq:3S18:LS2-CF-Li2} \\
    &&\hspace{-1cm}C^{(\text{Li}_2)}[{}^3S_1^{[8]},\text{LS-2},C_A]= \frac{2 (\tau -1) \left(\tau ^2+4 \tau +1\right)}{(\tau +1)^4}\text{Li}_2(-\tau ) \nonumber \\
    && -\frac{\left(3 \tau ^3+6 \tau ^2-3 \tau -2\right)}{(\tau +1)^4} \text{Li}_2(-2 \tau -1).\label{eq:3S18:LS2-CA-Li2}
\end{eqnarray}
While the remaining parts are:
\begin{eqnarray}
&& C^{(\ln)}[{}^3S_1^{[8]},\text{LS-1},C_F]= \frac{L_2 \left( [L_2-4 L_1^{(+)}] \left(\tau ^2+5 \tau -2\right)+4 \tau  \left(\tau ^2+8 \tau +7\right)\right)}{2 \tau  (\tau
   +1)^4} \nonumber \\
&& +\frac{1}{3 (\tau +1)^3 (2 \tau +1)^2} \Biggl\{-12 \ln(2) (\tau +1) (\tau  (\tau  (\tau +22)+19)+4) \nonumber \\
&& -(2 \tau +1) \left(6 (\tau +1)^3+\pi ^2 (\tau +5) (2 \tau -1) (2 \tau
   +1)\right) \nonumber \\
&& -12 (\tau +1) (\tau  (\tau  (\tau +22)+19)+4) \ln (\tau +1) \Biggr\}, \label{eq:3S18:LS1-CF-ln} \\
   && C^{(\ln)}[{}^3S_1^{[8]},\text{LS-1},C_A]=-\frac{L_2 \left( [L_2-4 L_1^{(+)}] \left(\tau ^2+5 \tau -2\right)+4 \tau  \left(\tau ^2+8 \tau +7\right)\right)}{2 \tau  (\tau
   +1)^4} \nonumber \\
   && +\frac{1}{6 (\tau +1)^5 (2 \tau +1)} \Biggl\{ 4 \ln(2) (\tau  (\tau  (\tau  (20 \tau +69)+137)+75)+11) (\tau +1) \nonumber \\
   && +4 (\tau +3) (2 \tau +1) (\tau +1)^3+\pi ^2 (\tau  (\tau  (\tau  (29-4 (\tau
   -5) \tau )+27)+3)-3) \nonumber \\
   && +(2 \tau +1) \ln (\tau ) (2 (\tau +1) (\tau  (\tau  (13 \tau +19)+35)+5) \nonumber \\ 
   &&-3 (\tau -1) (\tau  (\tau  (2 \tau +3)+6)+1) \ln
   (\tau )) \nonumber \\
   && +24 \tau  (\tau  (\tau  (11-(\tau -2) \tau )+13)+6) \ln (\tau +1)+24 \ln (\tau +1) \Biggr\}, \label{eq:3S18:LS1-CA-ln}
\end{eqnarray}
\begin{eqnarray}
 && C^{(\ln)}[{}^3S_1^{[8]},\text{LS-2},C_F]= -\frac{L_2 (L^{(+)}_1 (4-8 \tau )+L_2 (2 \tau -1)+12 \tau  (\tau +1))}{2 \tau  (\tau +1)^3} \nonumber \\
 && +\frac{2}{3 (\tau +1)^2 (2 \tau +1)} \Biggl\{ 6 \ln(2) (\tau +1) (5 \tau +2)+\pi ^2 \left(4 \tau ^2-1\right) \nonumber \\
 && +6 (\tau +1) (5 \tau +2) \ln (\tau +1)\Biggr\}, \label{eq:3S18:LS2-CF-ln} \\
    && C^{(\ln)}[{}^3S_1^{[8]},\text{LS-2},C_A]= \frac{L_2 (L^{(+)}_1 (4-8 \tau )+L_2 (2 \tau -1)+12 \tau  (\tau +1))}{2 \tau  (\tau +1)^3} \\
    && + \frac{1}{6 (\tau +1)^4 (2 \tau +1)} \Biggl\{-4 \ln(2) (\tau  (\tau  (25 \tau +76)+47)+8) (\tau +1) \nonumber \\
    && +2 (2 \tau +1) \left(\tau  \left((\tau +2) \tau ^2-\pi ^2 \left(\tau
   ^2+3\right)-2\right)-1\right) \nonumber \\
   && +3 (\tau -1) (2 \tau +1) (\tau  (\tau +4)+1) \ln^2(\tau )-12 (\tau  (5 \tau +4)+1) (\tau +1)^2 \ln (\tau +1) \nonumber \\
   &&-2 (2\tau +1) (\tau  (5 \tau +22)+5) (\tau +1) \ln (\tau ) \Biggr\}.\label{eq:3S18:LS2-CA-ln}
\end{eqnarray}

The corresponding coefficients in front of $n_F$ are:
\begin{eqnarray}
    C[{}^3S_1^{[8]},\text{LS-1},n_F] &=& \frac{
    2 (\tau +1)+(1-\tau) \ln (\tau )}{3 (\tau +1)^2}, \\
    C[{}^3S_1^{[8]},\text{LS-2},n_F] &=& -\frac{\tau +\ln (\tau )+1 
    }{3 (\tau +1)}.
\end{eqnarray}

\begin{figure}
    \centering
    \includegraphics[width=0.49\textwidth]{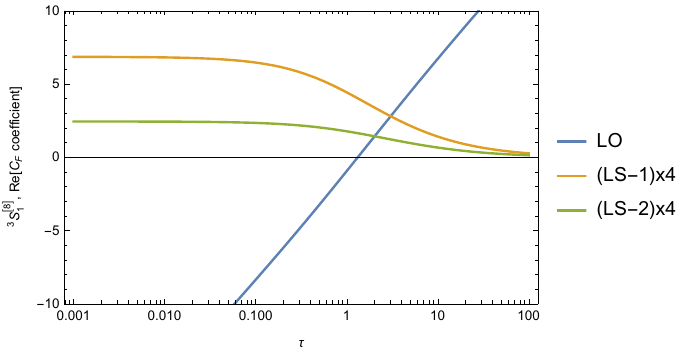}  \includegraphics[width=0.49\textwidth]{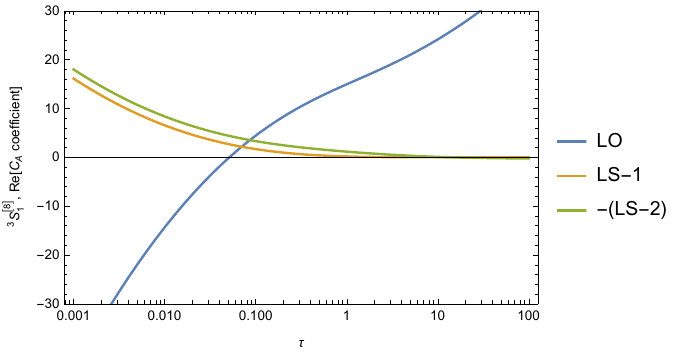}
    \caption{Plots of the coefficients in front of $C_F$ (left panel, Eqns.~(\ref{eq:3S18:LO-CF}), ((\ref{eq:3S18:LS1-CF-Li2})+(\ref{eq:3S18:LS1-CF-ln}))$\times 4$, ((\ref{eq:3S18:LS2-CF-Li2})+(\ref{eq:3S18:LS2-CF-ln}))$\times 4$ and $C_A$(right panel, Eqns.~(\ref{eq:3S18:LO-CA-Li2})+(\ref{eq:3S18:LO-CA-ln}), (\ref{eq:3S18:LS1-CA-Li2})+(\ref{eq:3S18:LS1-CA-ln}), $-$(\ref{eq:3S18:LS2-CA-Li2})$-$(\ref{eq:3S18:LS2-CA-ln}))}
    \label{fig:gR-3S18-CF-CA-plots}
\end{figure}

The plots of coefficients in front of the LO Lorentz structure (\ref{eq:LO-ampl_g+R->3S18}) and Lorentz structures (\ref{eq:gR-3S18:LS1}) and (\ref{eq:gR-3S18:LS2}) are shown in the Fig.~\ref{fig:gR-3S18-CF-CA-plots}. One notices that coefficients in front of new Lorentz structures have qualitatively similar behaviour with one-another, but very different from the behaviour of the coefficients in front of LO Lorentz structure. The asymptotics of finite parts of the coefficients in front of the LO Lorentz structure at $\tau \ll 1$ is:
\begin{eqnarray}
    &&\hspace{-5mm}C[{}^3S_1^{[8]},\text{LO},C_A]= -\ln ^2(\tau )+\ln (\tau )+\frac{2 \pi ^2}{3}+\frac{55}{9}-2 \ln (2) + O(\tau), \\ 
    &&\hspace{-5mm}C[{}^3S_1^{[8]},\text{LO},C_F]= 3 \ln (\tau )-\frac{3 \pi ^2}{2}-2+22 \ln (2) + O(\tau), \\
    &&\hspace{-5mm}C[{}^3S_1^{[8]},\text{LO},n_F]=-\ln (\tau ) -\frac{16}{9}+O(\tau). 
\end{eqnarray}

\begin{figure}
    \centering
    \includegraphics[width=0.49\textwidth]{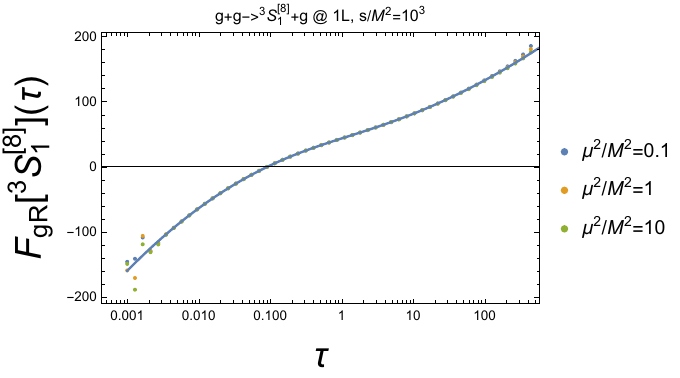}
    \caption{Comparison of the numerical results (points) for $F_{gR}[{}^3S_1^{[8]}]=C_F  C[{}^3S_1^{[8]},\text{LO},C_F] +  C_A C[{}^3S_1^{[8]},\text{LO},C_A]$, extracted from the exact one-loop amplitude of the process (\ref{test-proc:g+g-3S18}) for $s=10^3 M^2$ as described in the main text. Solid line -- analytic result of Eqns.~(\ref{eq:3S18:LO-CF})+(\ref{eq:3S18:LO-CA-Li2})+(\ref{eq:3S18:LO-CA-ln}). }
    \label{fig:gR-3S18-num-test}
\end{figure}

Finally, the same numerical cross-check as in the previous cases had been done, where the quantity $F_{gR}[{}^3S_1^{[8]}]=C_F  C[{}^3S_1^{[8]},\text{LO},C_F] +  C_A C[{}^3S_1^{[8]},\text{LO},C_A]$ was extracted from the numerical results for the interference of the one-loop corrected and tree-level amplitudes of the process:
\begin{equation}
    g(q) + g(k_1) \to Q\bar{Q}\left[ {}^3S_1^{[8]}\right](p) + g(k_2), \label{test-proc:g+g-3S18}
\end{equation}
with the help of Eqns.~(\ref{eq:QCD-EFT:gamma+g-1S08+g}) and (\ref{eq:g+R-1S08:CLO-expansion}), which allows one to test analytic results in Eqns.~(\ref{eq:3S18:LO-CF}) and (\ref{eq:3S18:LO-CA-Li2})+(\ref{eq:3S18:LO-CA-ln}). As it can be seen from the Fig.~\ref{fig:gR-3S18-num-test}, the numerical results (points) agree with the analytic expectation (solid line) within expected accuracy for a wide range of $\tau$, with the agreement being spoiled by numerical instability for $\tau< 0.003$ and by $O(-t/s)$ power corrections at very large $\tau>300$.

\section{Rapidity factorisation schemes}
\label{sec:rapidity-schemes}
In this section the relation of the results obtained above in the tilted-Wilson-line regularisation scheme to results in three other regularisation schemes for rapidity divergence which are used in high-energy QCD are given. The considered schemes are:
\begin{enumerate}
\item {\it The BFKL scheme} (Sec. \ref{sec:BFKL-scheme}), which is used for the calculation of pair production cross sections at large rapidity separation between components of the pair in the NLL approximation.
\item {\it The HEF scheme} (Sec. \ref{sec:HEF-scheme}), which is used for the computation of the HEF-resummed partonic cross section of quarkonium production at forward rapidities beyond DLA of Refs.~\cite{Lansberg:2021vie,Lansberg:2023kzf}. In this scheme the modes belonging to the impact-factor have {\it target} light-cone component of loop momentum $l_-<\Lambda_-$.  

\item {\it The shockwave scheme} (Sec. \ref{sec:Shockwave-scheme}) is relevant for the computation of the one-Reggeon-exchange part of the forward heavy quarkonium production cross section in shockwave picture of high-energy QCD~\cite{Balitsky:2001re, Balitsky:1995ub}.  In this approach, the regularisation of rapidity divergences is done by the cut of the {\it projectile} component of the loop momentum ($l_+$). The modes belonging to the impact-factor have $l_+>\Lambda_+$. The computation of genuine multi-Reggeon exchange contributions, which are also present in the saturation/shockwave approaches is beyond the scope of the present paper.
\end{enumerate}

\begin{figure}
    \centering
    \includegraphics[width=0.45\linewidth]{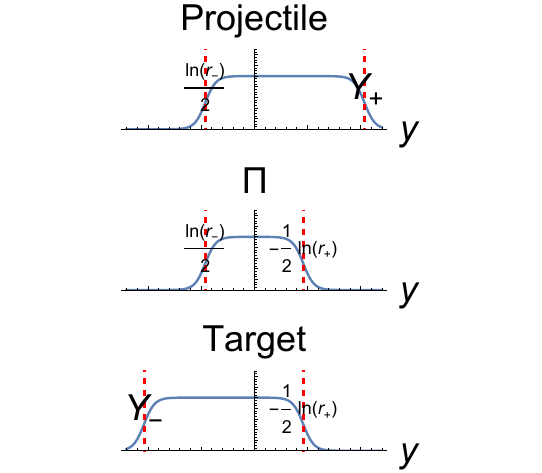}
    \includegraphics[width=0.45\linewidth]{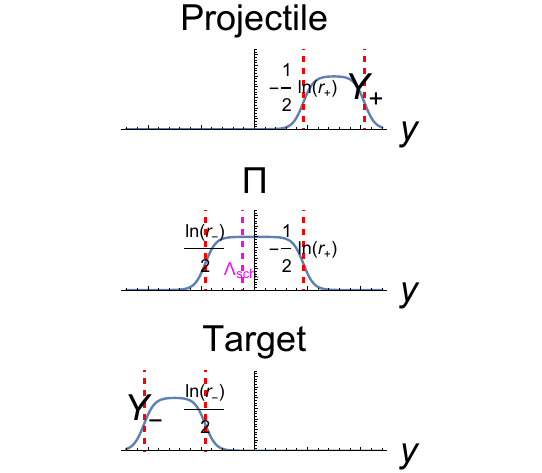}
    \caption{Schematic illustrations of the dependence of an integrands of one-loop corrections to the EFT quantities as functions of the rapidity ($y$) associated with the loop momentum. Upper plots correspond to the projectile impact-factor, middle plots to the Reggeon self-energy (\ref{eq:Pi-split}) and lower plots -- to the target impact-factor. Left-panel -- before localisation, right panel -- after localisation. }
    \label{fig:rap-local}
\end{figure}

In all cases, the computation of the transition term, which converts expressions obtained in the tilted-Wilson-line regularisation to any other scheme  amounts to the splitting of the term $-\Pi^{(1)}(t)$ in Eq.~(\ref{eq:QCD-EFT:gamma+g-1S08+g}) into contributions which belong to the ``target'' and ``projectile''/``impact-factor'' regions and subtracting
the latter contribution form the impact-factor result obtained in the tilted-Wilson-line scheme. 

This prescription is based on the procedure of localisation in rapidity for loop corrections in the EFT, which was proposed in Ref.~\cite{Chachamis:2012cc} at one loop and tested at two loops in Ref.~\cite{Chachamis:2013hma}. This procedure is illustrated schematically in the Fig.~\ref{fig:rap-local}. It is convenient for the sake of this explanatin to introduce different rapidity regulators for rapidity divergences in different directions: $r_{\pm}$, such that $\tilde{l}_+=l_+ + r_- l_-$ and $\tilde{l}_-=l_- + r_+ l_+$. Introducing the rapidity of the loop momentum $y=\ln (l_+/l_-)/2$ one obtains:
\begin{equation}
\frac{1}{\tilde{l}_{\pm}} = \frac{1}{l_\pm + r_{\mp} l_{\mp}} = \frac{1}{\sqrt{l^2+{\bf l}_T^2}} \frac{1}{e^{\pm y} + r_{\mp} e^{\mp y}},    
\end{equation}
from where one can see that the regularisation of rapidity divergences in tilted-Wilson scheme can be understood as a smooth cut in rapidity of the loop momentum. Regularising the denominator $1/l_+\to 1/\tilde{l}_+$ we constrain  $y>\ln (r_-)/2$ and regularising $1/l_-\to 1/\tilde{l}_-$ we require $y<-\ln (r_+)/2$. The same interpretation holds for rapidity divergences in real-emission diagrams. 

The rapidity integration in the projectile impact-factor is cut-off from above by some natural rapidity scale associated with a given process $Y_+$ (e.g. $Y_+\sim \ln (q_+/|{\bf p}_T|)$ in Fig.~\ref{fig:HEF-schematic}) and from below by $\ln (r_-)/2$. For the target impact-factor in the EFT, there is a natural rapidity-cut from below $Y_-$ (e.g. $Y_-\sim -\ln ((x_2 P_-)/|{\bf p}_T|)$ in Fig.~\ref{fig:HEF-schematic}) and from above the parton rapidity is cut by $-\ln (r_+)/2$, while in the one-loop correction to the Reggeon propagator there are both $1/\tilde{l}_+$ and $1/\tilde{l}_-$ denominators, so the rapidity is cut form both sides. From the left panel of the Fig.~\ref{fig:rap-local} one realises that the dependence on $r_{\pm}$ can not cancel between these contributions if they are just added together because the contribution of $\ln (r_-)/2<y<-\ln (r_+)/2$ will be double-counted.

However, if one {\it localises in rapidity} both impact-factors by subtracting from each of them the propagator contribution, then the dependence on $r_{\pm}$ will be depicted by the right panel of the Fig.~\ref{fig:rap-local} and summing the localised impact-factors with the propagator contribution will lead to cancellation of rapidity divergences. Since in this procedure one subtracts the Reggeon self-energy contribution twice and adds it back once, it is equivalent to just flipping the sign of the Reggeon self-energy contribution, which explains the minus sign in front of it in Eq.~(\ref{eq:QCD-EFT:gamma+g-1S08+g}). One can also observe, that the propagator correction which we {\it add} to the localised impact-factors (right panel of the Fig.~\ref{fig:rap-local}) can be split between projectile and target contributions according to any rapidity factorisation prescription, which is denoted in Fig.~\ref{fig:rap-local} by the cut $\Lambda_{\rm sch.}$. This is done for the case of HEF and shockwave schemes in Sec.~\ref{sec:HEF-scheme} and \ref{sec:Shockwave-scheme}, while transition to BFKL scheme (Sec.~\ref{sec:BFKL-scheme}) amounts to simple reshuffling of the Regge limit of the one-loop amplitude in terms of BFKL impact-factors and Reggeised gluon contributions.    

The one-loop correction to the Reggeised gluon propagator, which was computed for the first time in Refs.~\cite{Chachamis:2012cc,Chachamis:2012gh} and is given by the diagrams 11-13 in the Fig. 3 of the Ref.~\cite{Nefedov:2019mrg} is:
\begin{eqnarray}
&&    \Pi^{(1)}(-t,r_+,r_-) = \frac{\bar{\alpha}_s}{4\pi} \left\{ - \frac{(2-3d)C_A+2(d-2)n_F}{2(d-1)}B(-t) - C_A (-t) B_{[+-]}(-t) \right\} \nonumber \\
 &&= \Pi^{(1)}_{\text{non-RD}}(-t) - \frac{\bar{\alpha}_s C_A}{4\pi} (-t) B_{[+-]}(-t,r_+,r_-). \label{eq:Pi-split}
\end{eqnarray}
 In the Eq.~(\ref{eq:Pi-split}) the terms originating from non-rapidity-divergent diagrams are collected into into the $\Pi^{(1)}_{\text{(non-RD)}}$:
\begin{equation}
     \Pi^{(1)}_{\text{non-RD}}(-t) =  \frac{\bar{\alpha}_s}{4\pi} \left(\frac{\mu^2}{-t} \right)^{\epsilon} \left[ \frac{1}{\epsilon}(\beta_0-2C_A) - \frac{8}{3}C_A + \frac{5}{3}\beta_0 + O(\epsilon) \right],
\end{equation}
while the rapidity-divergent integral $B_{[+-]}$ is defined as~\cite{Nefedov:2019mrg}:
\begin{equation}
    B_{[+-]}({\bf k}_{T}^2,r_+,r_-) = \int \frac{[d^D l]}{l^2 (k+l)^2 [\tilde{l}_+] [\tilde{l}_-]}, \label{eq:B[+-]}
\end{equation}
where, and the regularisation parameter $r$ in impact-factors computed in the present paper can be identified with $r_-$ since it regularises poles in $l_+$. The result for $B_{[+-]}$ is~\cite{Nefedov:2019mrg}:
\begin{equation}
    B_{[+-]}({\bf k}_{T}^2,r_+,r_-) = \frac{1}{{\bf k}_T^2}\left(\frac{\mu^2}{{\bf k}_T^2} \right)^\epsilon \frac{\ln r_+ + \ln r_- + i\pi}{\epsilon} + O(r_+, r_-).
\end{equation}

\subsection{BFKL scheme}
\label{sec:BFKL-scheme}

Without loss of generality, one can refer to the Regge limit of the one-loop amplitude of a particular QCD process, e.g. process (\ref{test-proc:gamma+g-1S08}) to derive the relation between one-loop corrections to impact factors in the BFKL approach and one-loop corrections obtained in the high-energy EFT formalism~\cite{Lipatov95}. In the BFKL approach, the real part of one-loop scattering amplitude of the process (\ref{test-proc:gamma+g-1S08}) in the Regge limit is separated into contributions of a projectile ($\Gamma_{\text{proj}}$), a target ($\Gamma_g$) impact-factors\footnote{In case of amplitude (\ref{eq:QCD-EFT:gamma+g-1S08+g}), the target is a gluon} and of the gluon Regge trajectory ($\omega_g(-t)$), as follows:
\begin{equation}
\frac{\text{Re }{\cal M}^{\text{(1-loop)}}_{\text{proc. (\ref{test-proc:gamma+g-1S08})}}}{{\cal M}^{\text{(Born)}}_{\text{proc. (\ref{test-proc:gamma+g-1S08})}}} = \Gamma_{\text{proj}}(s_0,t) + \Gamma_g(s_0,t) + \omega_g(-t) \ln \frac{s}{s_0}, \label{eq:Regge-limit-BFKL}
\end{equation}
where $s=q_+ k_1^-$ in case of the process  (\ref{test-proc:gamma+g-1S08}), the one-loop Regge trajectory of a gluon is:
\begin{equation}
 \omega_g(-t) = \frac{\bar{\alpha}_s C_A}{2\pi} \frac{1}{\epsilon} \left(\frac{\mu^2}{-t} \right)^\epsilon,
\end{equation}
and the scale $s_0$ is the arbitrary rapidity factorisation scale, which is often chosen to be $s_0\sim -t$. Equating the expression (\ref{eq:Regge-limit-BFKL}) with the EFT expression (\ref{eq:QCD-EFT:gamma+g-1S08+g}) and separating symmetrically the non-rapidity-divergent part of $\Pi^{(1)}$ between projectile and target impact-factors, one obtains the following formulas for the BFKL impact-factors in terms of EFT quantities:
\begin{eqnarray}
 &&\hspace{-11mm} \Gamma_{\text{proj}}(s_0,t)= {
 \rm Re
 }\,C[\ldots+R\to\ldots] - \frac{1}{2} \Pi_{\text{non-RD}}^{(1)}(-t) +\frac{\bar{\alpha}_s C_A}{4\pi} \left(\frac{\mu^2}{-t} \right)^{\epsilon} \left[ \ln r - 2\ln \frac{q_+}{\sqrt{s_0}} \right], \label{eq:BFKL-sch} \\
&&\hspace{-11mm} \Gamma_{g}(s_0,t)= \gamma_{gRg}^{(1)}(k_1^-,-t) - \frac{1}{2} \Pi_{\text{non-RD}}^{(1)}(-t) +\frac{\bar{\alpha}_s C_A}{4\pi} \left(\frac{\mu^2}{-t} \right)^{\epsilon} \left[ \ln r - 2\ln \frac{k_1^-}{\sqrt{s_0}} \right].   
\end{eqnarray}

As one can see e.g. from the Eq.~(\ref{eq:gamma+R-1S08:CLO-expansion}), both the $\ln r$ and dependence on $q_+$ cancels in Eq.~(\ref{eq:BFKL-sch}), as it should be in the BFKL impact-factor, which is free from the large $\ln s$. Needless to say, that the Eq.~(\ref{eq:BFKL-sch}) is universal and any of the results (\ref{eq:gamma+R-1S08:CLO-expansion}), (\ref{eq:g+R-1S01:CLO-expansion}), (\ref{eq:g+R-1S08:CLO-expansion}) or (\ref{eq:g+R-3S18:CLO-expansion}) can be substituted to it.

\subsection{HEF scheme}
\label{sec:HEF-scheme}

To derive the transition term to the HEF-scheme, one inserts the
\begin{equation}
1=\theta(|l_-|>\Lambda_-) + \theta(|l_-|<\Lambda_-), \label{eq:unit-HEF} 
\end{equation}
into the integral (\ref{eq:B[+-]}) and attributes the contribution with $|l_-|>\Lambda_-$ {\it to the projectile impact-factor.} The reason is, that this cut will regularise the $1/l_-$-divergence in Eq.~(\ref{eq:B[+-]}), while the $1/l_+$-divergence still will be regularised by the $r_-=r$, so that the $r$-divergences will cancel when one, following the localisation in rapidity procedure (Fig.~\ref{fig:rap-local}), {\it subtracts} the $\Pi^{(1)}$ split by this cut (with $\Pi_{\text{non-RD}}^{(1)}$ being distributed equally between target and projectile) from the EFT result  for the projectile impact-factor. The integral (\ref{eq:B[+-]}) with this cut is:
\begin{equation}
     B^{(>)}_{[+-]}({\bf k}_{T}^2,r_-,\Lambda_-) = \int\frac{(\mu^2)^\epsilon d^d l}{i\pi^{d/2} r_\Gamma} \frac{ \theta(|l_-|>\Lambda_-)}{l_- [\tilde{l}_+][l^2+i\eta] [(k+l)^2+i\eta]},
\end{equation}
integrating-out the $l_+$ by the residue in the pole at $l_+=-r_-l_- + i\eta$, then integrating in $l_-$ from $\Lambda_-$ to $+\infty$ and expanding in $r_-\ll 1$ one obtains:
\begin{eqnarray}
  && B^{(>)}_{[+-]}({\bf k}_{T}^2,r_-,\Lambda_-) = \int\frac{ (\mu^2)^\epsilon d^{2-2\epsilon} {\bf l}_T}{\pi^{1-\epsilon} r_\Gamma}\left\{ -\frac{\ln r_- }{2{\bf l}_T^2 ({\bf l}_T+{\bf k}_T)^2} 
  \right. \nonumber \\
  && \left. + \frac{{\bf l}_T^2 \ln \frac{({\bf l}_T+{\bf k}_T)^2}{\Lambda_-^2} - ({\bf l}_T+{\bf k}_T)^2 \ln \frac{{\bf l}_T^2}{\Lambda_-^2}}{2 {\bf l}_T^2 ({\bf l}_T+{\bf k}_T)^2 [{\bf l}_T^2 - ({\bf l}_T+{\bf k}_T)^2]} \right\},\label{eq:B+-:gtr0}
\end{eqnarray}
which after integrating-out ${\bf l}_T$ gives:
\begin{eqnarray}
    B^{(>)}_{[+-]}({\bf k}_{T}^2,r_-,\Lambda_-) &=& \frac{1}{{\bf k}_T^2} \left(\frac{\mu^2}{{\bf k}_T^2} \right)^\epsilon \frac{1}{\epsilon} \left\{ \ln r_- +  \ln \frac{\Lambda_-^2}{{\bf k}_T^2} -\pi \cot (\pi \epsilon) + \gamma_E + \psi(-2\epsilon)  \right. \nonumber \\
&&  \left.  - \frac{1}{\epsilon} \left[ \frac{1}{2} + \epsilon \left( \pi\csc(2\pi\epsilon) + \psi(\epsilon) - \psi(2\epsilon)\right)\right] \right\} \nonumber \\
  &=&  \frac{1}{{\bf k}_T^2} \left(\frac{\mu^2}{{\bf k}_T^2} \right)^\epsilon \left\{ -\frac{1}{\epsilon^2} + \frac{1}{\epsilon}\left[ \ln r_- + \ln \frac{\Lambda_-^2}{{\bf k}_T^2} \right] - \frac{\pi^2}{6}  +O(\epsilon) \right\}. \label{eq:Bpm_HEF}
\end{eqnarray}

With this result at hands, one can relate the one-loop impact-factors in the tilted-Wilson-line scheme and the HEF scheme as:
\begin{eqnarray}
 &&   {\rm Re }\,C_{\rm HEF}[\ldots+R_-({\bf k}_T)\to\ldots] = {\rm Re }\,C[\ldots+R_-({\bf k}_T)\to\ldots] - \frac{1}{2} \Pi_{\text{non-RD}}^{(1)}({\bf k}_T^2) \nonumber \\
 && + \frac{\bar{\alpha}_s C_A}{4\pi} {\bf k}_T^2 B^{(>)}_{[+-]}({\bf k}_{T}^2,r,\Lambda_-),
\end{eqnarray}
where by the ${\rm Re }\,C[\ldots+R_-({\bf k}_T)\to\ldots]$ in this formula one understands the {\it non-localised} correction to the impact-factor in the EFT, i.e. Eqns. (\ref{eq:gamma+R-1S08:CLO-expansion}), (\ref{eq:g+R-1S01:CLO-expansion}), (\ref{eq:g+R-1S08:CLO-expansion}) or (\ref{eq:g+R-3S18:CLO-expansion}) as they stand.

\subsection{Shockwave scheme}
\label{sec:Shockwave-scheme}

To obtain the transition term for the projectile impact-factor in the shockwave scheme, one splits the integral (\ref{eq:B[+-]}) according to the value of the {\it projectile} light-cone component of the loop momentum:
\begin{equation}
1=\theta(|l_+|>\Lambda_+) + \theta(|l_+|<\Lambda_+), \label{eq:unit-Shockwave} 
\end{equation}
but this time, one attributes the contribution with $\theta(|l_+|<\Lambda_+)$ to the projectile impact-factor, because it will regularise the denominator $1/l_-$ in the integral (\ref{eq:B[+-]}), while the denominator $1/l_+$ will be still regularised by $r_-=r$. {\it Subtracting} the $\Pi^{(1)}$ split by this cut (with $\Pi_{\text{non-RD}}^{(1)}$ being distributed equally between target and projectile) from the EFT impact-factor in tilted-Wilson-line scheme one cancels the $\ln r_-$-divergence and is left with $\Lambda_+$-dependent impact-factor in the shackwave scheme.   The integral (\ref{eq:B[+-]}) with the cut is:
\begin{equation}
     B^{(<)}_{[+-]}({\bf k}_{T}^2,r_-,\Lambda_+) = \int\frac{(\mu^2)^\epsilon d^d l}{i\pi^{d/2} r_\Gamma} \frac{ \theta(|l_+|<\Lambda_+)}{[l_-] [\tilde{l}_+][l^2+i\eta] [(k+l)^2+i\eta]}.
\end{equation}
For $0>l_+>\Lambda_+$, one integrates-out the $l_-$ by residues in poles $l_-=+i\eta$ and $l_-=-l_+/r+i\eta$ of the eikonal propagators and obtains (the case  $-\Lambda_+<l_+<0$ gives the overall factor of 2):
\begin{eqnarray}
  &&B^{(<)}_{[+-]}({\bf k}_{T}^2,r_-,\Lambda_+) = \int\frac{(\mu^2)^\epsilon d^{2-2\epsilon} {\bf l}_T}{\pi^{1-\epsilon} r_\Gamma} \int\limits_0^{\Lambda_+} \frac{dl_+}{l_+} \nonumber \\
  &&\times \left[ \frac{1}{{\bf l}_T^2({\bf k}_T+{\bf l}_T)^2} - \frac{r_-^2}{[l_+^2 + r_-{\bf l}_T^2 ] [l_+^2 + r_-({\bf k}_T+{\bf l}_T)^2]} \right]  \nonumber \\
  && = \int\frac{(\mu^2)^\epsilon d^{2-2\epsilon} {\bf l}_T}{\pi^{1-\epsilon} r_\Gamma}\left\{ -\frac{\ln r_- }{2{\bf l}_T^2 ({\bf l}_T+{\bf k}_T)^2} 
   - \frac{{\bf l}_T^2 \ln \frac{({\bf l}_T+{\bf k}_T)^2}{\Lambda_+^2} - ({\bf l}_T+{\bf k}_T)^2 \ln \frac{{\bf l}_T^2}{\Lambda_+^2}}{2 {\bf l}_T^2 ({\bf l}_T+{\bf k}_T)^2 [{\bf l}_T^2 - ({\bf l}_T+{\bf k}_T)^2]} \right\},
\end{eqnarray}
note the change of the sign of the second term in curly brackets, compared with Eq.~(\ref{eq:B+-:gtr0}). The result for this integral is:
\begin{eqnarray}
    B^{(<)}_{[+-]}({\bf k}_{T}^2,r_-,\Lambda_+) &=& \frac{1}{{\bf k}_T^2} \left(\frac{\mu^2}{{\bf k}_T^2} \right)^\epsilon \frac{1}{\epsilon} \left\{ \ln r_- -  \ln \frac{\Lambda_+^2}{{\bf k}_T^2} +\pi \cot (\pi \epsilon) - \gamma_E - \psi(-2\epsilon)  \right. \nonumber \\
&&  \left.  + \frac{1}{\epsilon} \left[ \frac{1}{2} + \epsilon \left( \pi\csc(2\pi\epsilon) + \psi(\epsilon) - \psi(2\epsilon)\right)\right] \right\} \nonumber \\
  &=&  \frac{1}{{\bf k}_T^2} \left(\frac{\mu^2}{{\bf k}_T^2} \right)^\epsilon \left\{ \frac{1}{\epsilon^2} + \frac{1}{\epsilon}\left[ \ln r_- - \ln \frac{\Lambda_+^2}{{\bf k}_T^2} \right] + \frac{\pi^2}{6}  +O(\epsilon) \right\}, \label{eq:Bpm_Shockwave}
\end{eqnarray}
and the impact-factor in the shockwave scheme is related with the impact-factor in the tilted-Wilson-line regularisation as:
\begin{eqnarray}
 &&   {\rm Re }\,C_{\text{shockw.}}[\ldots+R_-({\bf k}_T)\to\ldots] = {\rm Re }\,C[\ldots+R_-({\bf k}_T)\to\ldots] - \frac{1}{2} \Pi_{\text{non-RD}}^{(1)}({\bf k}_T^2) \nonumber \\
 && + \frac{\bar{\alpha}_s C_A}{4\pi} {\bf k}_T^2 B^{(<)}_{[+-]}({\bf k}_{T}^2,r,\Lambda_+),
\end{eqnarray}
where by the ${\rm Re }\,C[\ldots+R_-({\bf k}_T)\to\ldots]$ in this formula one understands the {\it non-localised} correction to the impact-factor in the EFT, i.e. Eqns. (\ref{eq:gamma+R-1S08:CLO-expansion}), (\ref{eq:g+R-1S01:CLO-expansion}), (\ref{eq:g+R-1S08:CLO-expansion}) or (\ref{eq:g+R-3S18:CLO-expansion}) as they stand.

\section{Conclusions and outlook}
\label{sec:conclusions}

In the present paper the results for one-loop corrections to impact-factors of gluon or photon-induced production of $Q\bar{Q}$-pairs in ${}^1S_0^{[1]}$, ${}^1S_0^{[8]}$ and  ${}^3S_1^{[8]}$ states via interaction with a Reggeised gluon ($R$) had been obtained. Taken together with corresponding real-emission corrections these results can be used to compute full NLO impact-factors which in turn can be employed to provide the NLL-resummed HEF predictions for the production of heavy quarkonia at forward rapidities. Also these results can be used to provide BFKL-resummed results in the NLL approximation for heavy-quarkonium pair production, or production of quarkonium+jet or quarkonium+light-meson pair, at high separation in rapidity between the components of the pair. One can also envision the application of obtained results for the study of forward production of heavy quarkonia in the dilute-dense scattering, taking into account gluon saturation effects through the shockwave formalism. However, to this end the genuine multi-Reggeon exchange contributions should be consistently added to the results obtained in the present paper. 

Another fruitful direction of future development is the computation of {\it central} production vertices of the type $R_+R_-\to Q\bar{Q}[{}^{2S+1}L_J^{[1,8]}]$ at one loop. Such computations are likely to not to require computation of any new scalar integrals and are essentially a reduction problem. The EFT formalism~\cite{Lipatov95} employed in the present paper is essentially the only known formalism in low-$x$ physics which allows one to do such computations in practice. 

\appendix

\section{Deriving one-loop quarkonium production amplitudes with \texttt{FormCalc}\label{appendix:FormCalc}}

\subsection{Tensor reduction with \texttt{FormCalc} and \texttt{LoopTools}}
The chain of tools combining \texttt{FeynArts}~\cite{FeynArts}, \texttt{FormCalc}~\cite{Hahn:2000jm} and library of one-loop integrals \texttt{LoopTools}~\cite{Hahn:1998yk}, provides a flexible framework for Feynman-diagram-based one-loop computations in Standard Model and beyond, especially if one is only after numerical results for the amplitude. However the computation of heavy quarkonium production amplitudes is not implemented there by default. Fortunately, the above-mentioned chain of tools is sufficiently flexible to allow one to derive tree-level and one-loop production amplitudes with $S$-wave states of $Q\bar{Q}$ in the final state without the need for any modifications of the packages themselves. For example, in order to derive one of the amplitudes for the processes (\ref{test-proc:g+g-1S01}), (\ref{test-proc:g+g-1S08}) or (\ref{test-proc:g+g-3S18})\footnote{For the process (\ref{test-proc:gamma+g-1S08}) one just replaces $g(k_1)\to \gamma(k_1)$ in Eq.~(\ref{App:proc:gg-QQg}) which does not require any significant change in the described procedure} one starts by generating expressions for all one loop amplitudes corresponding to Feynman diagrams contributing to the process: 
\begin{equation}
g(k_1) + g (k_2) \to Q(p_1) + \bar{Q}(p_2) + g(k_4), \label{App:proc:gg-QQg}
\end{equation}
using \texttt{FeynArts}. These expressions can be easily manipulated in Mathematica. To project-out a desired state of the $Q\bar{Q}$-pair, the heavy quark spinors in obtained expressions are replaced with the corresponding projectors (\ref{eq:SpinProj-0}) or (\ref{eq:SpinProj-1}) and colour indices of heavy quarks are contracted with  projectors on the colour-singlet or colour-octet state, mentioned in the Sec.~\ref{sec:LO}. Finally, momenta of heavy quarks are substituted as $p_{1,2}^\mu \to p^\mu/2$, to project-out the $L=0$ state. After these substitutions, the amplitudes of $2\to 3$ process (\ref{App:proc:gg-QQg}) effectively turn into amplitudes of the desired $2\to 2$ process (\ref{test-proc:g+g-1S01}), (\ref{test-proc:g+g-1S08}) or (\ref{test-proc:g+g-3S18}), and corresponding technical information, accompanying expressions generated by \texttt{FeynArts} should be changed in order to facilitate the correct processing of these expressions by \texttt{FORMCalc}. 

The next step is to take the interference between one-loop and tree-level diagrams of the desired process, perform summations over polarisations and colours of initial- and final-state particles between the amplitude and complex-conjugate amplitude and to perform the colour, Dirac algebra and Lorentz index contractions in the numerator of the amplitude. All these steps are done by \texttt{FORMCalc} in a fully automated way, once correctly-formatted expressions for the amplitudes are provided from Mathematica. As a result one obtains an interference term between any Born and loop corrected diagram of the process under consideration in a form of the following (Lorentz-)scalar quantity:
\begin{equation}
    {\cal D}_i=\int \frac{d^d l}{(2\pi)^d} \frac{{\rm Num}(l^2,l\cdot k_1,l\cdot k_2,l\cdot k_3)}{ D^{(i)}_1 \ldots D^{(i)}_{n_i}}, \label{App:eq:Di}
\end{equation}
where the numerator ${\rm Num}(l^2,l\cdot k_1,l\cdot k_2,l\cdot k_3)$ is a function of $l^2$ and scalar products of independent external momenta $k_{1,2,3}$ with the loop momentum $l$ as well as of scalar products $k_i\cdot k_j$, Feynman's denominators are denoted as $D^{(i)}_j=(q_j^{(i)}+l)^2-(m_j^{(i)})^2+i\eta$ with $q_j^{(i)}$ being linear combinations of $k_{1,2,3}$ and $m_j^{(i)}=m_Q$ or 0 for the case under consideration. If the number of linearly-independent denominators $n_i\leq 4$ then all scalar products with the loop momentum can be expressed in terms of denominators un-ambiguously and \texttt{FormCalc} can perform the reduction of ${\cal D}_i$ down to scalar integrals\footnote{More precisely, \texttt{FormCalc} does the first step of the Passarino-Veltman reduction, reducing the amplitude down to tensor-integral coefficients, and reduction formulas for those, in terms of usual one-loop scalar integrals, are implemented in \texttt{LoopTools}. So the linear dependence of denominators mentioned above actually becomes a problem on on the level of \texttt{LoopTools}, where it leads to vanishing of some Gram-determinants in the reduction formulas. }.

However, as mentioned in Sec.~\ref{sec:reduction}, due to the momentum assignment $p_{1,2}^\mu \to p^\mu/2$ for heavy quarks, some denominators become linearly-dependent in certain diagrams. This will lead to the failure of tensor reduction, so the linear dependence of denominators has to be resolved {\it before} exporting expressions for Feynman diagrams from Mathematica to \texttt{FORMCalc}. It is straightforward to get rid of linear dependence, using the partial-fractioning procedure described in Sec.~\ref{sec:reduction} around Eq.~(\ref{eq:lin-rel-Di}). For those diagrams, for which the free term $a\neq 0$ in Eq.~(\ref{eq:lin-rel-Di}), no denominator gets squared after partial fractioning, and resulting expressions can be directly fed into \texttt{FORMCalc}. However the case when $a=0$ requires special treatment, because one of the denominators is squared while there is no integrals with squared denominators in the standard basis of scalar one-loop integrals implemented in \texttt{LoopTools}. One possible way to deal with this problem, which was used to obtain numerical results in the present paper, is described in the next subsection.  

\subsection{Dealing with Coulomb-divergent diagrams}

In the present calculation, the degenerate case with $a=0$ in Eq.~(\ref{eq:lin-rel-Di}) is realised only in those diagrams where the outgoing heavy quark lines are connected by the gluon line, closing the loop. Diagrams from this class contain the Coulomb divergence, which is automatically put to zero by dimensional regularisation, as discussed in Sec.~\ref{sec:reduction}. In such diagrams, the outgoing quark propagators and a gluon propagator which connects them are linearly dependent and the corresponding diagrams can be put into the following general form:
\begin{equation}
    {\cal D}_{\rm Coul.}=\int\frac{d^D l}{(2\pi)^D} \frac{F(l)}{[l^2+i\eta] [(p_Q+l)^2-m_Q^2+i\eta] [(p_Q-l)^2-m_Q^2+i\eta]} , \label{eq:D-Coul}
\end{equation}
where $p_Q=p/2$, $p_Q^2=m_Q^2=M^2/4$ and the function $F(l)$ encapsulates the whole numerator of the amplitude ${\rm Num(l)}$ and all other propagators besides those which are written explicitly in Eq.~(\ref{eq:D-Coul}). The linear dependence of denominators in the diagram can be resolved as follows: 
\begin{eqnarray}
{\cal D}_{\rm Coul.}&=& \int\frac{d^D l}{(2\pi)^D} \frac{1}{[l^2+i\eta]^2 [(p_Q+l)^2-m_Q^2+i\eta]} \frac{F(l) + F(-l)}{2},\label{eq:Coul-PF}
\end{eqnarray}
where the gluon propagator got squared. To get rid of the square one would be tempted to use the quantity:
\begin{eqnarray}
 \overline{\cal D}_{\rm Coul.}(\lambda) &=&  \frac{1}{\lambda^2} \left\{ \int\frac{d^D l}{(2\pi)^D} \frac{1}{[l^2-\lambda^2+i\eta] [(p_Q+l)^2-m_Q^2+i\eta]} \frac{F(l) + F(-l)}{2} \right. \nonumber \\
 &-& \left. \int\frac{d^D l}{(2\pi)^D} \frac{1}{[l^2+i\eta] [(p_Q+l)^2-m_Q^2+i\eta]} \frac{F(l) + F(-l)}{2}  \right\}, \label{eq:Coul-PF-lam}
\end{eqnarray}
which is equal to ${\cal D}_{\rm Coul.}$ in the limit $\lambda\to 0$ at finite $\epsilon$ and has an advantage that in each term of the Eq.~(\ref{eq:Coul-PF-lam}) all denominators are linearly-independent and raised to the first power, so in principle this quantity can be straightforwardly computed by \texttt{FORMCalc}+\texttt{LoopTools} at finite $\lambda$. The catch of course is, that the limit $\lambda\to 0$ does not commute with expansion in $\epsilon$, so the quantity $\overline{\cal D}_{\rm Coul.}$, computed at finite $\lambda$ will have different structure of $1/\epsilon$ poles than the desired quantity ${\cal D}_{\rm Coul.}$. These two objects are related identically, for any $\lambda$, as follows:
\begin{equation}
 {\cal D}_{\rm Coul.} = \overline{\cal D}_{\rm Coul.}(\lambda) + {\cal E}(\lambda, F), \label{eq:Coul-PF-decomp}   
\end{equation}
where:
\begin{eqnarray}
 {\cal E}(\lambda, F) &=& \int\frac{d^D l}{(2\pi)^D} \frac{1}{[l^2+i\epsilon] [(p_Q+l)^2-m_Q^2+i\epsilon] }\left[ \frac{1}{l^2+i\epsilon} - \frac{1}{l^2-\lambda^2+i\epsilon} \right] \frac{F(l) + F(-l)}{2} \nonumber \\
 &=& - \lambda^2 \int\frac{d^D l}{(2\pi)^D} \frac{1}{[l^2+i\eta]^2[l^2-\lambda^2+i\eta][(p_Q+l)^2-m_Q^2+i\eta]} \frac{F(l) + F(-l)}{2} . \label{eq:Coul-err-term} 
\end{eqnarray}
One can see, that the contribution of $l^2 \gg \lambda^2$ into Eq.~(\ref{eq:Coul-err-term}) is suppressed by the factor in square brackets, so it is likely that the behaviour of $ {\cal E}(\lambda, F)$ for small $\lambda$ can be estimated in a simple way in terms of $F(l\simeq 0)$. In fact, the following expansion for the Eq.~(\ref{eq:Coul-err-term}) holds up to $O(\lambda^2)$:
\begin{eqnarray}
  {\cal E}(\lambda,F) &=& J(\lambda,m_Q,\epsilon) \left[ F(0) + \frac{\lambda^2}{2(D-1)} \left(g^{\mu\nu} - \frac{p_Q^{\mu} p_Q^{\nu}}{m_Q^2} \right) \left. \frac{\partial^2 F(l)}{\partial l_{\mu} \partial l_{\nu}} \right\vert_{l=0}  \right] + O(\lambda^2) \label{eq:theo1-lam}
\end{eqnarray}
where the integral
\begin{eqnarray}
&& J(\lambda,m_Q,\epsilon) = - \lambda^2 \int\frac{d^D l}{(2\pi)^D} \frac{1}{[l^2+i\eta]^2[l^2-\lambda^2+i\eta][(p+l)^2-m_Q^2+i\eta]} \label{eq:theo1:J-def} \\
 &&= \frac{i\pi^2}{(2\pi)^4} \frac{(4\pi)^\epsilon}{\Gamma(1-\epsilon)} \frac{m_Q^{-2-2\epsilon}}{2} \left[ \frac{2\pi m_Q}{\lambda} + \ln \frac{\lambda^2}{m_Q^2} - \frac{1}{\epsilon} -\frac{\pi\lambda}{4m_Q} + O(\epsilon,\lambda^2) \right].\label{eq:theo1:J} 
\end{eqnarray}
A sketch of the proof of the expansion (\ref{eq:theo1-lam}) is given in the next section. The quantity $F(0)$ or even the derivative of $F(l)$ at $l=0$ does not contain any integration and therefore is straightforward to compute, and for sufficiently small $\lambda$ the $\lambda$-dependence should cancel in the Eq.~(\ref{eq:Coul-PF-decomp}). 

The downside of this procedure is that the integral (\ref{eq:theo1:J}) behaves as $1/\lambda$ for $\lambda\to 0$, so large numerical cancellations are happening in Eq.~(\ref{eq:Coul-PF-decomp}), which may lead to the numerical instability. But in practice this problem turns out to be not very severe and working with $\lambda/m_Q\sim 10^{-4}$ and using the Mathematica interface of \texttt{LoopTools} as provider of one-loop Feynman integrals, one obtains  $O(0.1\%)$ precision for the quantities plotted with points in Figs.~\ref{fig:gamma+g-1S08_plots}, \ref{fig:g+g-1S01_plots}, \ref{fig:g+g-1S08_plots} and \ref{fig:gR-3S18-num-test} over wide range of $\tau$. Since this calculation is done in a way completely independent form the procedure used for the calculations in the EFT, the agreement between analytic and numerical results shown in Figs.~\ref{fig:gamma+g-1S08_plots}, \ref{fig:g+g-1S01_plots}, \ref{fig:g+g-1S08_plots} and \ref{fig:gR-3S18-num-test} serves as a strong cross-check of both methods.

\subsection{Expansion of the error term}
To study the error term (\ref{eq:Coul-err-term}) one expands the function $(F(l)+F(-l))/2$ in a Taylor series around $l=0$:
\[
\frac{F(l)+F(-l)}{2}= F(0) + \left. \frac{\partial^2 F(l)}{\partial l_{\mu_1} \partial l_{\mu_2}} \right\vert_{l=0} \frac{l_{\mu_1}l_{\mu_2}}{2!}  + \left. \frac{\partial^4 F(l)}{\partial l_{\mu_1} \partial l_{\mu_2} \partial l_{\mu_3} \partial l_{\mu_4}}\right\vert_{l=0} \frac{l_{\mu_1}l_{\mu_2}l_{\mu_3}l_{\mu_4}}{4!} + O(l^6).
\]

Substituting this expansion into the Eq.~(\ref{eq:Coul-err-term}) one can see, that  the expansion (\ref{eq:theo1-lam}) will be proven if it turns-out, that tensor integrals of the type:
\[
J_{\mu_1\ldots \mu_n} = - \lambda^2 \int\frac{d^D l}{(2\pi)^D} \frac{l_{\mu_1}\ldots l_{\mu_n}}{(l^2)^2[l^2-\lambda^2][(p_Q+l)^2-m_Q^2]} 
\]
all go to zero sufficiently rapidly in the limit $\lambda\to 0$. 

Let's consider the simplest case $J_{\mu_1 \mu_2}$. This tensor integral can be expressed as:
\begin{equation}
J^{\mu_1 \mu_2} = \frac{1}{D-1} \left[ \left( g^{\mu_1\mu_2} - \frac{p_Q^{\mu_1} p_Q^{\mu_2}}{m_Q^2}\right) J_g + \left( D \frac{p_Q^{\mu_1} p_Q^{\mu_2}}{m_Q^2} - g_{\mu_1 \mu_2} \right) \frac{J_{pp}}{m_Q^2}\right],\label{eq:Jmn-PaVe}
\end{equation}
where:
\begin{eqnarray}
 J_{g}=g^{\mu_1 \mu_2} J_{\mu_1 \mu_2},\ J_{pp}=p_Q^{\mu_1} p_Q^{\mu_2}J_{\mu_1\mu_2}.
\end{eqnarray}

Let's first consider:
\begin{eqnarray}
&& J_g = - \lambda^2 \int\frac{d^D l}{(2\pi)^D} \frac{(l^2-\lambda^2) + \lambda^2 }{(l^2)^2[l^2-\lambda^2][(p_Q+l)^2-m_Q^2]} \nonumber \\
&&= -\underbrace{\lambda^2  \int\frac{d^D l}{(2\pi)^D} \frac{1}{(l^2)^2[(p_Q+l)^2-m_Q^2]}}_{O(\lambda^2)} +\lambda^2 \underbrace{J(\lambda,m_Q,\epsilon)}_{\sim 1/\lambda} = O(\lambda), \label{eq:int-est}
\end{eqnarray}
so this integral goes to zero in the limit $\lambda\to 0$. 

For the second case
\[
J_{pp}= - \lambda^2 \int\frac{d^D l}{(2\pi)^D} \frac{ (p_Ql)^2 }{(l^2)^2[l^2-\lambda^2][(p_Q+l)^2-m_Q^2]}
\]
one have to use
\[
(p_Ql)=\frac{1}{2}\left[ [(p_Q+l)^2-m_Q^2] - [l^2-\lambda^2] - \lambda^2 \right],
\]
to obtain
\begin{eqnarray*}
 &&J_{pp}= - \frac{\lambda^2}{2} \underbrace{ \int\frac{d^D l}{(2\pi)^D} \frac{ (p_Ql) }{(l^2)^2[l^2-\lambda^2]} }_{=0}  + \underbrace{\frac{\lambda^2}{2} \int\frac{d^D l}{(2\pi)^D} \frac{ (p_Ql) }{(l^2)^2 [(p_Q+l)^2-m_Q^2]}}_{O(\lambda^2)}  \\
 &&+ \frac{\lambda^4}{2} \int\frac{d^D l}{(2\pi)^D} \frac{ (p_Ql) }{(l^2)^2[l^2-\lambda^2][(p_Q+l)^2-m_Q^2]} \\ 
 && = O(\lambda^2) + \frac{\lambda^4}{2} \int\frac{d^D l}{(2\pi)^D} \frac{ (pl) }{(l^2)^2[l^2-\lambda^2][(p_Q+l)^2-m_Q^2]}  \\
 &&= O(\lambda^2) + \underbrace{\frac{\lambda^4}{4} \underbrace{ \int\frac{d^D l}{(2\pi)^D} \frac{ 1 }{(l^2)^2[l^2-\lambda^2]} }_{\sim 1/\lambda^2}}_{O(\lambda^2)} - \underbrace{\frac{\lambda^4}{4} \int\frac{d^D l}{(2\pi)^D} \frac{ 1 }{(l^2)^2 [(p_Q+l)^2-m_Q^2]}}_{O(\lambda^4)} \\
 && -\underbrace{\frac{\lambda^6}{4} J(\lambda,m_Q,\epsilon)}_{O(\lambda^5)}=O(\lambda^2).  
\end{eqnarray*}
so the integral $p_Q^{\mu_1} p_Q^{\mu_2}J^{\mu_1\mu_2}$ also goes to zero, hence the tensor integral $J_{\mu_1 \mu_2}$ goes to zero as well.

As shown above, the only $O(\lambda)$ contribution comes from $J_g=\lambda^2 J(\lambda,m_Q,\epsilon)+O(\lambda^2)$. Substituting it to the Eq. (\ref{eq:Jmn-PaVe}) one obtains the $O(\lambda)$ correction term in Eq.~(\ref{eq:theo1-lam}).

The case of $J_{\mu_1\mu_2\mu_3\mu_4}$ can be considered using the same arguments and one finds that this tensor integral is suppressed as $O(\lambda^2)$. This pattern of suppression continues for the integrals with higher tensor rank, hence the $O(\lambda^2)$ estimation of the remainder term of the expansion in Eq.~(\ref{eq:theo1-lam}).

\acknowledgments

Author would like to thank Thomas Hahn, Jean-Philippe Lansberg and Vladyslav Shtabovenko for useful discussions.

This project has received funding from the European Union's Horizon 2020
research and innovation programme under grant agreement No.~101065263
for the Marie Sk{\l}odowska-Curie action ``RadCor4HEF'', under grant
agreement No.~824093 in order to contribute to the EU Virtual Access
{\sc NLOAccess} and to the JRA Fixed-Target Experiments at the LHC.
 This project has also received funding from the Agence Nationale de la
Recherche (ANR) via the grant ANR-20-CE31-0015 (``PrecisOnium'') and via
the IDEX Paris-Saclay ``Investissements d’Avenir'' (ANR-11-IDEX-0003-01)
through the GLUODYNAMICS project funded by the ``P2IO LabEx (ANR-10-LABX-0038)''.
This work  was also partly supported by the French CNRS via the IN2P3 projects ``GLUE@NLO'' and ``QCDFactorisation@NLO'' as well as via the COPIN-IN2P3 project \#12-147 ``$k_T$ factorisation and quarkonium production in the LHC era''.


\bibliographystyle{JHEP}
\bibliography{mybibfile}

\providecommand{\href}[2]{#2}\begingroup\raggedright\begin{thebibliography}{10}

\bibitem{Brambilla:2010cs}
N.~Brambilla et~al., \emph{{Heavy Quarkonium: Progress, Puzzles, and Opportunities}}, \href{https://doi.org/10.1140/epjc/s10052-010-1534-9}{\emph{Eur. Phys. J. C} {\bfseries 71} (2011) 1534} [\href{https://arxiv.org/abs/1010.5827}{{\ttfamily 1010.5827}}].

\bibitem{Brambilla:2014jmp}
N.~Brambilla et~al., \emph{{QCD and Strongly Coupled Gauge Theories: Challenges and Perspectives}}, \href{https://doi.org/10.1140/epjc/s10052-014-2981-5}{\emph{Eur. Phys. J. C} {\bfseries 74} (2014) 2981} [\href{https://arxiv.org/abs/1404.3723}{{\ttfamily 1404.3723}}].

\bibitem{Lansberg:2019adr}
J.-P.~Lansberg, \emph{{New Observables in Inclusive Production of Quarkonia}}, \href{https://doi.org/10.1016/j.physrep.2020.08.007}{\emph{Phys. Rept.} {\bfseries 889} (2020) 1} [\href{https://arxiv.org/abs/1903.09185}{{\ttfamily 1903.09185}}].

\bibitem{Arbuzov:2020cqg}
A.~Arbuzov et~al., \emph{{On the physics potential to study the gluon content of proton and deuteron at NICA SPD}}, \href{https://doi.org/10.1016/j.ppnp.2021.103858}{\emph{Prog. Part. Nucl. Phys.} {\bfseries 119} (2021) 103858} [\href{https://arxiv.org/abs/2011.15005}{{\ttfamily 2011.15005}}].

\bibitem{Chapon:2020heu}
E.~Chapon et~al., \emph{{Perspectives for quarkonium studies at the high-luminosity LHC}},  \href{https://arxiv.org/abs/2012.14161}{{\ttfamily 2012.14161}}.

\bibitem{Bodwin:1994jh}
G.T.~Bodwin, E.~Braaten and G.P.~Lepage, \emph{{Rigorous QCD analysis of inclusive annihilation and production of heavy quarkonium}}, \href{https://doi.org/10.1103/PhysRevD.55.5853}{\emph{Phys. Rev. D} {\bfseries 51} (1995) 1125} [\href{https://arxiv.org/abs/hep-ph/9407339}{{\ttfamily hep-ph/9407339}}].

\bibitem{Lansberg:2021vie}
J.-P.~Lansberg, M.~Nefedov and M.A.~Ozcelik, \emph{{Matching next-to-leading-order and high-energy-resummed calculations of heavy-quarkonium-hadroproduction cross sections}}, \href{https://doi.org/10.1007/JHEP05(2022)083}{\emph{JHEP} {\bfseries 05} (2022) 083} [\href{https://arxiv.org/abs/2112.06789}{{\ttfamily 2112.06789}}].

\bibitem{Lansberg:2023kzf}
J.-P.~Lansberg, M.~Nefedov and M.A.~Ozcelik, \emph{{Curing the high-energy perturbative instability of vector-quarkonium-photoproduction cross sections at order $\alpha \alpha _s^3$ with high-energy factorisation}}, \href{https://doi.org/10.1140/epjc/s10052-024-12588-x}{\emph{Eur. Phys. J. C} {\bfseries 84} (2024) 351} [\href{https://arxiv.org/abs/2306.02425}{{\ttfamily 2306.02425}}].

\bibitem{Catani:1990xk}
S.~Catani, M.~Ciafaloni and F.~Hautmann, \emph{{GLUON CONTRIBUTIONS TO SMALL x HEAVY FLAVOR PRODUCTION}}, \href{https://doi.org/10.1016/0370-2693(90)91601-7}{\emph{Phys. Lett. B} {\bfseries 242} (1990) 97}.

\bibitem{Catani:1990eg}
S.~Catani, M.~Ciafaloni and F.~Hautmann, \emph{{High-energy factorization and small x heavy flavor production}}, \href{https://doi.org/10.1016/0550-3213(91)90055-3}{\emph{Nucl. Phys. B} {\bfseries 366} (1991) 135}.

\bibitem{Collins:1991ty}
J.C.~Collins and R.K.~Ellis, \emph{{Heavy quark production in very high-energy hadron collisions}}, \href{https://doi.org/10.1016/0550-3213(91)90288-9}{\emph{Nucl. Phys.} {\bfseries B360} (1991) 3}.

\bibitem{Catani:1994sq}
S.~Catani and F.~Hautmann, \emph{{High-energy factorization and small x deep inelastic scattering beyond leading order}}, \href{https://doi.org/10.1016/0550-3213(94)90636-X}{\emph{Nucl. Phys.} {\bfseries B427} (1994) 475} [\href{https://arxiv.org/abs/hep-ph/9405388}{{\ttfamily hep-ph/9405388}}].

\bibitem{DelDuca:2001gu}
V.~Del~Duca and E.W.N.~Glover, \emph{{The High-energy limit of {QCD} at two loops}}, \href{https://doi.org/10.1088/1126-6708/2001/10/035}{\emph{JHEP} {\bfseries 10} (2001) 035} [\href{https://arxiv.org/abs/hep-ph/0109028}{{\ttfamily hep-ph/0109028}}].

\bibitem{Fadin:2017nka}
V.S.~Fadin and L.N.~Lipatov, \emph{{Reggeon cuts in {QCD} amplitudes with negative signature}}, \href{https://doi.org/10.1140/epjc/s10052-018-5910-1}{\emph{Eur. Phys. J.} {\bfseries C78} (2018) 439} [\href{https://arxiv.org/abs/1712.09805}{{\ttfamily 1712.09805}}].

\bibitem{Caron-Huot:2017fxr}
S.~Caron-Huot, E.~Gardi and L.~Vernazza, \emph{{Two-parton scattering in the high-energy limit}}, \href{https://doi.org/10.1007/JHEP06(2017)016}{\emph{JHEP} {\bfseries 06} (2017) 016} [\href{https://arxiv.org/abs/1701.05241}{{\ttfamily 1701.05241}}].

\bibitem{BFKL1}
E.A.~Kuraev, L.N.~Lipatov and V.S.~Fadin, \emph{Multi - {Reggeon} processes in the {Yang-Mills} theory}, {\emph{Sov. Phys. JETP} {\bfseries 44} (1976) 443}.

\bibitem{BFKL2}
E.A.~Kuraev, L.N.~Lipatov and V.S.~Fadin, \emph{The {Pomeranchuk} singularity in {non-Abelian} gauge theories}, {\emph{Sov. Phys. JETP} {\bfseries 45} (1977) 199}.

\bibitem{BFKL3}
Y.Y.~Balitsky and L.N.~Lipatov, \emph{The {Pomeranchuk} singularity in {Quantum Chromodynamics}}, {\emph{Sov. J. Nucl. Phys.} {\bfseries 28} (1978) 822}.

\bibitem{He:di-Jpsi}
Z.-G.~He, B.A.~Kniehl, M.A.~Nefedov and V.A.~Saleev, \emph{{Double Prompt $J/\psi$ Hadroproduction in the Parton Reggeization Approach with High-Energy Resummation}}, \href{https://doi.org/10.1103/PhysRevLett.123.162002}{\emph{Phys. Rev. Lett.} {\bfseries 123} (2019) 162002} [\href{https://arxiv.org/abs/1906.08979}{{\ttfamily 1906.08979}}].

\bibitem{Celiberto:2022dyf}
F.G.~Celiberto and M.~Fucilla, \emph{{Diffractive semi-hard production of a $J/\psi $ or a $\Upsilon $ from single-parton fragmentation plus a jet in hybrid factorization}}, \href{https://doi.org/10.1140/epjc/s10052-022-10818-8}{\emph{Eur. Phys. J. C} {\bfseries 82} (2022) 929} [\href{https://arxiv.org/abs/2202.12227}{{\ttfamily 2202.12227}}].

\bibitem{Lipatov95}
L.N.~Lipatov, \emph{Gauge invariant effective action for high-energy processes in {QCD}}, \href{https://doi.org/10.1016/0550-3213(95)00390-E}{\emph{Nucl. Phys.} {\bfseries B452} (1995) 369}.

\bibitem{Caron-Huot:2013fea}
S.~Caron-Huot, \emph{{When does the gluon reggeize?}}, \href{https://doi.org/10.1007/JHEP05(2015)093}{\emph{JHEP} {\bfseries 05} (2015) 093} [\href{https://arxiv.org/abs/1309.6521}{{\ttfamily 1309.6521}}].

\bibitem{Hahn:2000jm}
T.~Hahn, \emph{{Automatic loop calculations with FeynArts, FormCalc, and LoopTools}}, \href{https://doi.org/10.1016/S0920-5632(00)00848-3}{\emph{Nucl. Phys. B Proc. Suppl.} {\bfseries 89} (2000) 231} [\href{https://arxiv.org/abs/hep-ph/0005029}{{\ttfamily hep-ph/0005029}}].

\bibitem{MH_PolePrescr}
M.~Hentschinski, \emph{{Pole prescription of higher order induced vertices in Lipatov's {QCD} effective action}}, \href{https://doi.org/10.1016/j.nuclphysb.2012.02.001}{\emph{Nucl. Phys.} {\bfseries B859} (2012) 129}.

\bibitem{RevLipatov97}
L.N.~Lipatov, \emph{Small x physics in perturbative {QCD}}, \href{https://doi.org/10.1016/S0370-1573(96)00045-2}{\emph{Phys. Rept.} {\bfseries 286} (1997) 131}.

\bibitem{BondZubkov}
S.~Bondarenko and M.A.~Zubkov, \emph{{The dimensionally reduced description of the high energy scattering and the effective action for the reggeized gluons}}, \href{https://doi.org/10.1140/epjc/s10052-018-6089-1}{\emph{Eur. Phys. J.} {\bfseries C78} (2018) 617}.

\bibitem{Chachamis:2012cc}
G.~Chachamis, M.~Hentschinski, J.D.~Madrigal~Martinez and A.~Sabio~Vera, \emph{{Next-to-leading order corrections to the gluon-induced forward jet vertex from the high energy effective action}}, \href{https://doi.org/10.1103/PhysRevD.87.076009}{\emph{Phys. Rev.} {\bfseries D87} (2013) 076009}.

\bibitem{Chachamis:2012gh}
G.~Chachamis, M.~Hentschinski, J.D.~Madrigal~Martinez and A.~Sabio~Vera, \emph{{Quark contribution to the gluon Regge trajectory at {NLO} from the high energy effective action}}, \href{https://doi.org/10.1016/j.nuclphysb.2012.03.015}{\emph{Nucl. Phys.} {\bfseries B861} (2012) 133}.

\bibitem{Chachamis:2013hma}
G.~Chachamis, M.~Hentschinski, J.D.~Madrigal~Martinez and A.~Sabio~Vera, \emph{{Gluon Regge trajectory at two loops from Lipatov's high energy effective action}}, \href{https://doi.org/10.1016/j.nuclphysb.2013.08.013}{\emph{Nucl. Phys.} {\bfseries B876} (2013) 453}.

\bibitem{Nefedov:2019mrg}
M.A.~Nefedov, \emph{{Computing one-loop corrections to effective vertices with two scales in the EFT for Multi-Regge processes in QCD}}, \href{https://doi.org/10.1016/j.nuclphysb.2019.114715}{\emph{Nucl. Phys.} {\bfseries B946} (2019) 114715} [\href{https://arxiv.org/abs/1902.11030}{{\ttfamily 1902.11030}}].

\bibitem{Mangano:1996kg}
M.L.~Mangano and A.~Petrelli, \emph{{NLO quarkonium production in hadronic collisions}}, \href{https://doi.org/10.1142/S0217751X97002048}{\emph{Int. J. Mod. Phys. A} {\bfseries 12} (1997) 3887} [\href{https://arxiv.org/abs/hep-ph/9610364}{{\ttfamily hep-ph/9610364}}].

\bibitem{Belusca-Maito:2020ala}
H.~B\'elusca-Ma\"\i{}to, A.~Ilakovac, M.~Ma\dj{}or-Bo\v{z}inovi\'c and D.~St\"ockinger, \emph{{Dimensional regularization and Breitenlohner-Maison/\textquoteright{}t Hooft-Veltman scheme for $\gamma_5$ applied to chiral YM theories: full one-loop counterterm and RGE structure}}, \href{https://doi.org/10.1007/JHEP08(2020)024}{\emph{JHEP} {\bfseries 08} (2020) 024} [\href{https://arxiv.org/abs/2004.14398}{{\ttfamily 2004.14398}}].

\bibitem{Kniehl:2006sk}
B.A.~Kniehl, D.V.~Vasin and V.A.~Saleev, \emph{{Charmonium production at high energy in the $k_{T}$ -factorization approach}}, \href{https://doi.org/10.1103/PhysRevD.73.074022}{\emph{Phys. Rev. D} {\bfseries 73} (2006) 074022} [\href{https://arxiv.org/abs/hep-ph/0602179}{{\ttfamily hep-ph/0602179}}].

\bibitem{FeynArts}
T.~Hahn, \emph{{Generating Feynman diagrams and amplitudes with FeynArts 3}}, \href{https://doi.org/10.1016/S0010-4655(01)00290-9}{\emph{Comput. Phys. Commun.} {\bfseries 140} (2001) 418}.

\bibitem{FeynCalc}
R.~Mertig, M.~Bohm and A.~Denner, \emph{{{FEYN CALC:} Computer algebraic calculation of Feynman amplitudes}}, \href{https://doi.org/10.1016/0010-4655(91)90130-D}{\emph{Comput. Phys. Commun.} {\bfseries 64} (1991) 345}.

\bibitem{Shtabovenko:2016sxi}
V.~Shtabovenko, R.~Mertig and F.~Orellana, \emph{{New Developments in FeynCalc 9.0}}, \href{https://doi.org/10.1016/j.cpc.2016.06.008}{\emph{Comput. Phys. Commun.} {\bfseries 207} (2016) 432} [\href{https://arxiv.org/abs/1601.01167}{{\ttfamily 1601.01167}}].

\bibitem{Shtabovenko:2020gxv}
V.~Shtabovenko, R.~Mertig and F.~Orellana, \emph{{FeynCalc 9.3: New features and improvements}}, \href{https://doi.org/10.1016/j.cpc.2020.107478}{\emph{Comput. Phys. Commun.} {\bfseries 256} (2020) 107478} [\href{https://arxiv.org/abs/2001.04407}{{\ttfamily 2001.04407}}].

\bibitem{Chetyrkin:1981qh}
K.G.~Chetyrkin and F.V.~Tkachov, \emph{{Integration by Parts: The Algorithm to Calculate beta Functions in 4 Loops}}, \href{https://doi.org/10.1016/0550-3213(81)90199-1}{\emph{Nucl. Phys. B} {\bfseries 192} (1981) 159}.

\bibitem{Smirnov:2019qkx}
A.V.~Smirnov and F.S.~Chuharev, \emph{{FIRE6: Feynman Integral REduction with Modular Arithmetic}}, \href{https://doi.org/10.1016/j.cpc.2019.106877}{\emph{Comput. Phys. Commun.} {\bfseries 247} (2020) 106877} [\href{https://arxiv.org/abs/1901.07808}{{\ttfamily 1901.07808}}].

\bibitem{Shtabovenko:2016whf}
V.~Shtabovenko, \emph{{FeynHelpers: Connecting FeynCalc to FIRE and Package-X}}, \href{https://doi.org/10.1016/j.cpc.2017.04.014}{\emph{Comput. Phys. Commun.} {\bfseries 218} (2017) 48} [\href{https://arxiv.org/abs/1611.06793}{{\ttfamily 1611.06793}}].

\bibitem{Petrelli:1997ge}
A.~Petrelli, M.~Cacciari, M.~Greco, F.~Maltoni and M.L.~Mangano, \emph{{NLO production and decay of quarkonium}}, \href{https://doi.org/10.1016/S0550-3213(97)00801-8}{\emph{Nucl. Phys. B} {\bfseries 514} (1998) 245} [\href{https://arxiv.org/abs/hep-ph/9707223}{{\ttfamily hep-ph/9707223}}].

\bibitem{Hentschinski:2011tz}
M.~Hentschinski and A.~Sabio~Vera, \emph{{{NLO} jet vertex from Lipatov's {QCD} effective action}}, \href{https://doi.org/10.1103/PhysRevD.85.056006}{\emph{Phys. Rev.} {\bfseries D85} (2012) 056006}.

\bibitem{Patel:2015tea}
H.H.~Patel, \emph{{Package-X: A Mathematica package for the analytic calculation of one-loop integrals}}, \href{https://doi.org/10.1016/j.cpc.2015.08.017}{\emph{Comput. Phys. Commun.} {\bfseries 197} (2015) 276} [\href{https://arxiv.org/abs/1503.01469}{{\ttfamily 1503.01469}}].

\bibitem{Smirnov}
V.A.~Smirnov, \emph{Feynman Integral Calculus}, Springer-Verlag, Berlin, Heidelberg (2006).

\bibitem{Huber:2005yg}
T.~Huber and D.~Maitre, \emph{{HypExp: A Mathematica package for expanding hypergeometric functions around integer-valued parameters}}, \href{https://doi.org/10.1016/j.cpc.2006.01.007}{\emph{Comput. Phys. Commun.} {\bfseries 175} (2006) 122} [\href{https://arxiv.org/abs/hep-ph/0507094}{{\ttfamily hep-ph/0507094}}].

\bibitem{Kotikov:1990kg}
A.V.~Kotikov, \emph{{Differential equations method: New technique for massive Feynman diagrams calculation}}, \href{https://doi.org/10.1016/0370-2693(91)90413-K}{\emph{Phys. Lett. B} {\bfseries 254} (1991) 158}.

\bibitem{Nefedov:2023uen}
M.~Nefedov, \emph{{On the high-energy instability of quarkonium production}}, \href{https://doi.org/10.1016/j.nuclphysbps.2023.11.012}{\emph{Nucl. Part. Phys. Proc.} {\bfseries 343} (2024) 11} [\href{https://arxiv.org/abs/2309.09608}{{\ttfamily 2309.09608}}].

\bibitem{Nefedov:2020ecb}
M.~Nefedov, \emph{{Towards stability of NLO corrections in High-Energy Factorization via Modified Multi-Regge Kinematics approximation}}, \href{https://doi.org/10.1007/JHEP08(2020)055}{\emph{JHEP} {\bfseries 08} (2020) 055} [\href{https://arxiv.org/abs/2003.02194}{{\ttfamily 2003.02194}}].

\bibitem{Nefedov:2019mws}
M.~Nefedov, \emph{{One-loop corrections to multiscale effective vertices in the EFT for Multi-Regge processes in QCD}},  in \emph{{27th International Workshop on Deep Inelastic Scattering and Related Subjects (DIS 2019) Torino, Italy, April 8-12, 2019}}, 2019 [\href{https://arxiv.org/abs/1905.01105}{{\ttfamily 1905.01105}}].

\bibitem{Fadin:1992zt}
V.S.~Fadin and R.~Fiore, \emph{{Quark contribution to the gluon-gluon - reggeon vertex in QCD}}, \href{https://doi.org/10.1016/0370-2693(92)90696-2}{\emph{Phys. Lett. B} {\bfseries 294} (1992) 286}.

\bibitem{Fadin:1993wh}
V.S.~Fadin and L.N.~Lipatov, \emph{{Radiative corrections to QCD scattering amplitudes in a multi - Regge kinematics}}, \href{https://doi.org/10.1016/0550-3213(93)90168-O}{\emph{Nucl. Phys. B} {\bfseries 406} (1993) 259}.

\bibitem{DelDuca:1998kx}
V.~Del~Duca and C.R.~Schmidt, \emph{{Virtual next-to-leading corrections to the impact factors in the high-energy limit}}, \href{https://doi.org/10.1103/PhysRevD.57.4069}{\emph{Phys. Rev. D} {\bfseries 57} (1998) 4069} [\href{https://arxiv.org/abs/hep-ph/9711309}{{\ttfamily hep-ph/9711309}}].

\bibitem{Fadin:1999de}
V.S.~Fadin, R.~Fiore, M.I.~Kotsky and A.~Papa, \emph{{The Gluon impact factors}}, \href{https://doi.org/10.1103/PhysRevD.61.094005}{\emph{Phys. Rev. D} {\bfseries 61} (2000) 094005} [\href{https://arxiv.org/abs/hep-ph/9908264}{{\ttfamily hep-ph/9908264}}].

\bibitem{Balitsky:2001re}
I.~Balitsky, \emph{{Effective field theory for the small x evolution}}, \href{https://doi.org/10.1016/S0370-2693(01)01041-3}{\emph{Phys. Lett. B} {\bfseries 518} (2001) 235} [\href{https://arxiv.org/abs/hep-ph/0105334}{{\ttfamily hep-ph/0105334}}].

\bibitem{Balitsky:1995ub}
I.~Balitsky, \emph{{Operator expansion for high-energy scattering}}, \href{https://doi.org/10.1016/0550-3213(95)00638-9}{\emph{Nucl. Phys. B} {\bfseries 463} (1996) 99} [\href{https://arxiv.org/abs/hep-ph/9509348}{{\ttfamily hep-ph/9509348}}].

\bibitem{Hahn:1998yk}
T.~Hahn and M.~Perez-Victoria, \emph{{Automatized one loop calculations in four-dimensions and D-dimensions}}, \href{https://doi.org/10.1016/S0010-4655(98)00173-8}{\emph{Comput. Phys. Commun.} {\bfseries 118} (1999) 153} [\href{https://arxiv.org/abs/hep-ph/9807565}{{\ttfamily hep-ph/9807565}}].

\end{thebibliography}\endgroup
\end{document}